\documentclass[aps,prl,reprint,superscriptaddress,nolongbibliography]{revtex4-2}

\usepackage{graphicx,siunitx,xcolor}
\usepackage[italicdiff]{physics}

\usepackage{hyperref}
\hypersetup{
 colorlinks,
 linkcolor={blue!50!black},
 citecolor={blue!50!black},
 urlcolor={blue!80!black}
}
\DeclareMathOperator*{\argmax}{\arg\!\max}

\begin{document}
\title{Low-Frequency Vibrational States in Ideal Glasses with Random Pinning}
\author{Kumpei Shiraishi}
\email{kumpeishiraishi@g.ecc.u-tokyo.ac.jp}
\author{Yusuke Hara}
\author{Hideyuki Mizuno}
\affiliation{Graduate School of Arts and Sciences, University of Tokyo, Komaba, Tokyo 153-8902, Japan}
\date{\today}

\begin{abstract}
Glasses exhibit spatially localized vibrations in the low-frequency regime.
These localized modes emerge below the boson peak frequency $\omega_\text{BP}$, and their vibrational densities of state follow $g(\omega) \propto \omega^4$~($\omega$ is frequency).
Here, we attempt to address how the localized vibrations behave through the ideal glass transition.
To do this, we employ a random pinning method, which enables us to study the thermodynamic glass transition.
We find that the localized vibrations survive even in equilibrium glass states.
Remarkably, the localized vibrations still maintain the properties of appearance below $\omega_\text{BP}$ and $g(\omega) \propto \omega^4$.
Our results provide important insight into the material properties of ideal glasses.
\end{abstract}

\maketitle

\begin{figure*}
\centering
\includegraphics[width=\linewidth]{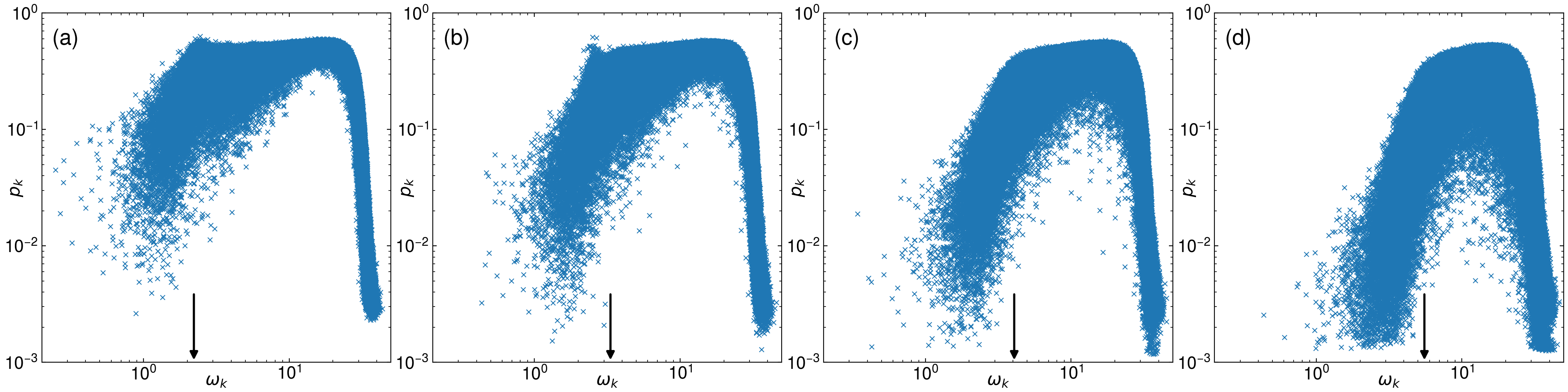}
\caption{
Participation ratio $p_k$ versus eigenfrequency $\omega_k$.
Fractions of pinned particles are (a) $c = 0.00$, (b) 0.05, (c) 0.10, and (d) 0.20.
Present data are constructed from 1000 configurations, each of which is composed of $N = 1000$ particles.
Arrows indicate values of the boson peak frequency $\omega_\text{BP}$.}
\label{fig:PR}
\end{figure*}

Recent progress has been made in our understanding of low-frequency vibration in glasses.
Mean-field theories such as the effective medium theory (EMT)~\cite{DeGiuli_2014} and the replica theory~\cite{Franz_2015} state that the vibrational density of states (VDOS) $g(\omega)$ follows the non-Debye scaling law of $g(\omega) \propto \omega^2$, which is different from the Debye law of $g(\omega) \propto \omega^{d-1}$ of crystals ($d$ denotes the spatial dimension)~\cite{Ashcroft_Mermin}.
Numerical simulations verified this theoretical prediction of $g(\omega) \propto \omega^2$ in the large-dimension limit~\cite{Charbonneau_2016,Shimada_2020}.
On the other hand, simulations of finite-dimensional glasses indicated that scaling of the VDOS $g(\omega) \propto \omega^4$ emerges even in the low-frequency regime~\cite{Lerner_2016,Mizuno_2017,Shimada_LJ_2018}.
This $\omega^4$ scaling occurs due to the contribution of spatially localized modes.
Most recently, theoretical works~\cite{Shimada:2020ka,Shimada:2020vi,Shimada:2020vc} successfully explained this scaling law in the framework of the EMT.

The localized modes in glasses have been intensively studied in recent years.
First, these modes have a spatial structure in which the strongly vibrating unstable core is surrounded by an energetically stable far-field region~\cite{Lerner_2016,Shimada_spatial_2018}.
Therefore, the modes are referred to as ``quasi''-localized vibrations.
Second, simulations of a three-dimensional polydisperse system indicated that the coefficient $A_4$ of the $\omega^4$ scaling, i.e., $g(\omega) = A_4 \omega^4$, significantly decreases as the temperature of equilibrium configurations is lowered~\cite{Wang_low_freq_2019,Rainone_PNAS_2020,Ji_2020}.
The study used the swap Monte Carlo (MC) method~\cite{Ninarello_2017} to anneal the system down to temperatures far below the mode-coupling temperature~\cite{Wang_low_freq_2019}.
If we extrapolate this result, it might be expected that the number of localized modes further decreases and even vanishes as the temperature is lowered toward the so-called ideal glass transition temperature.
In this work, we attempt to address how the localized vibrations behave through the thermodynamic (ideal) glass transition.

Proper sampling of glass configurations at low temperatures is challenging.
At low temperatures, particularly below the mode-coupling temperature $T_c$, the relaxation time dramatically increases and exceeds the realistic computational time with ordinary molecular dynamics (MD) or MC simulations.
Even when employing sampling techniques such as the replica-exchange method~\cite{Hukushima_1996,Yamamoto_2000}, it is difficult to sample configurations at temperatures far below $T_c$~\cite{DeMichele2002,Coslovich_EPJE_2018}.
Even if the system is tailored to the state-of-the-art swap MC~\cite{Grigera_2001,Gutierrez_2015,Ninarello_2017}, the accessible temperature is limited to approximately the ``experimental'' glass transition temperature $T_g$~\cite{Ninarello_2017}, which is much higher than the ``thermodynamic'' (ideal) glass transition temperature that we focus on in the present work.

The method of random pinning can realize the equilibrium glass states.
It has been shown theoretically in the mean-field framework that freezing a finite fraction of particles' positions can shift the thermodynamic glass transition to a relatively high temperature near $T_c$~\cite{Cammarota_2012}.
In numerical simulations of three-dimensional glass formers, thermodynamically equilibrium glass configurations were successfully realized with vanishing configurational entropy~\cite{Ozawa_2015} and distinctive overlap statistics of the thermodynamic glass transition~\cite{Kob_2013,Ozawa_2015}.
Using the random pinning technique, dynamics have been studied not only in the supercooled liquid regime~\cite{Kim_2003,Kim_2011,Jack2013,Kob_2014,Chakrabarty_2015,Chakrabarty2016} but also in the equilibrium glass states~\cite{Ozawa_PRL_2018}.
Interestingly, it was reported that transitions between different basins are induced by localized excitations even in equilibrium glasses.
Moreover, experimental implementation of random pinning was successfully achieved by optical tweezers for colloidal glasses~\cite{Gokhale_2014}.

Vibrational states of randomly pinned systems were studied by Angelani and colleagues~\cite{Angelani_2018}.
They revealed the $\omega^4$ scaling law of the VDOS, but up to \SI{90}{\percent} of the particles were pinned in their systems.
In this case, only a small number of unpinned particles are distributed among the chunk of pinned particles, so it is doubtful that these systems are reasonable for studying vibrational eigenmodes in solid-states.
In addition, much more important, the researchers focused on relatively high temperatures $T = 3T_c$ where the thermodynamic glass transition never occurs and only a crossover takes place~\cite{Cammarota_2012,Kob_2013,Ozawa_2015}.
Thus, it still remains to be addressed whether the localized vibrations can survive in thermodynamically equilibrium glasses.

Here, we deal with a lower temperature around $T_c$ and generate the equilibrium glass states.
We investigate the low-frequency vibrational properties of ideal glasses.
The random pinning method has the advantage of suppressing phonons and solving hybridizations between localized vibrations and phonons~\cite{Angelani_2018}.
We can therefore focus directly on the localized modes.
Remarkably, we find that localized vibrations and the boson peak (BP, excess low-frequency modes) survive even in ideal glasses.
In particular, localized vibrations always emerge below the BP frequency $\omega_\text{BP}$ and maintain the scaling law of $g(\omega) \propto \omega^4$ through the thermodynamic glass transition.
Our findings provide important insight into the material properties of ideal glasses.

We examine a standard model of amorphous systems: the Kob-Andersen (KA) model~\cite{Kob_Andersen_I_1995}.
We carry out MD simulations to prepare an equilibrium (supercooled) liquid state of $N$ particles at temperature $T_p$.
Here, we set $T_p$ to 0.45, which is close to the mode-coupling temperature $T_c = 0.435$~\cite{Kob_Andersen_PRL_1994}.
At this temperature, equilibrium glass states are clearly observed for our model~\cite{Ozawa_2015}.
Next, we randomly choose $cN$ particles where $c$ is in a range of $[0, 1]$ and permanently freeze those $cN$ particles.
Thus, $N_\text{up} = N - cN$ unpinned particles can move in the system.
Note that the system maintains the equilibrium state and does \textit{not} going into the nonequilibrium state through the random pinning operation~\cite{Cammarota_2012}.
As $c$ increases from zero, the system undergoes the thermodynamic glass transition at $c \approx 0.10$, at which the configurational entropy vanishes~\cite{Ozawa_2015}.

At each value of $c$, we quench the system to the inherent structure by minimizing the system potential, where we displace $N_\text{up}$ unpinned particles while keeping frozen $cN$ pinned particles.
For the minimization, we use the FIRE algorithm~\cite{Guenole_2020}.
We finally perform vibrational mode analysis to obtain eigenvalues $\lambda_k$ and eigenvectors $\vb*{e}_k = \pqty{\vb*{e}_k^1, \dots, \vb*{e}_k^{N_\text{up}}}$ for each eigenmode $k = 1, 2, \dots, 3N_\text{up}$.
The eigenfrequency $\omega_k$ is determined as $\omega_k = \sqrt{\lambda_k}$.
Note that $N_\text{up}$ unpinned particles participate in the vibrations, while $cN$ pinned particles are always frozen with no vibrations.
See the Supplemental Material for more details.

\begin{figure}
\centering
\includegraphics[width=\linewidth]{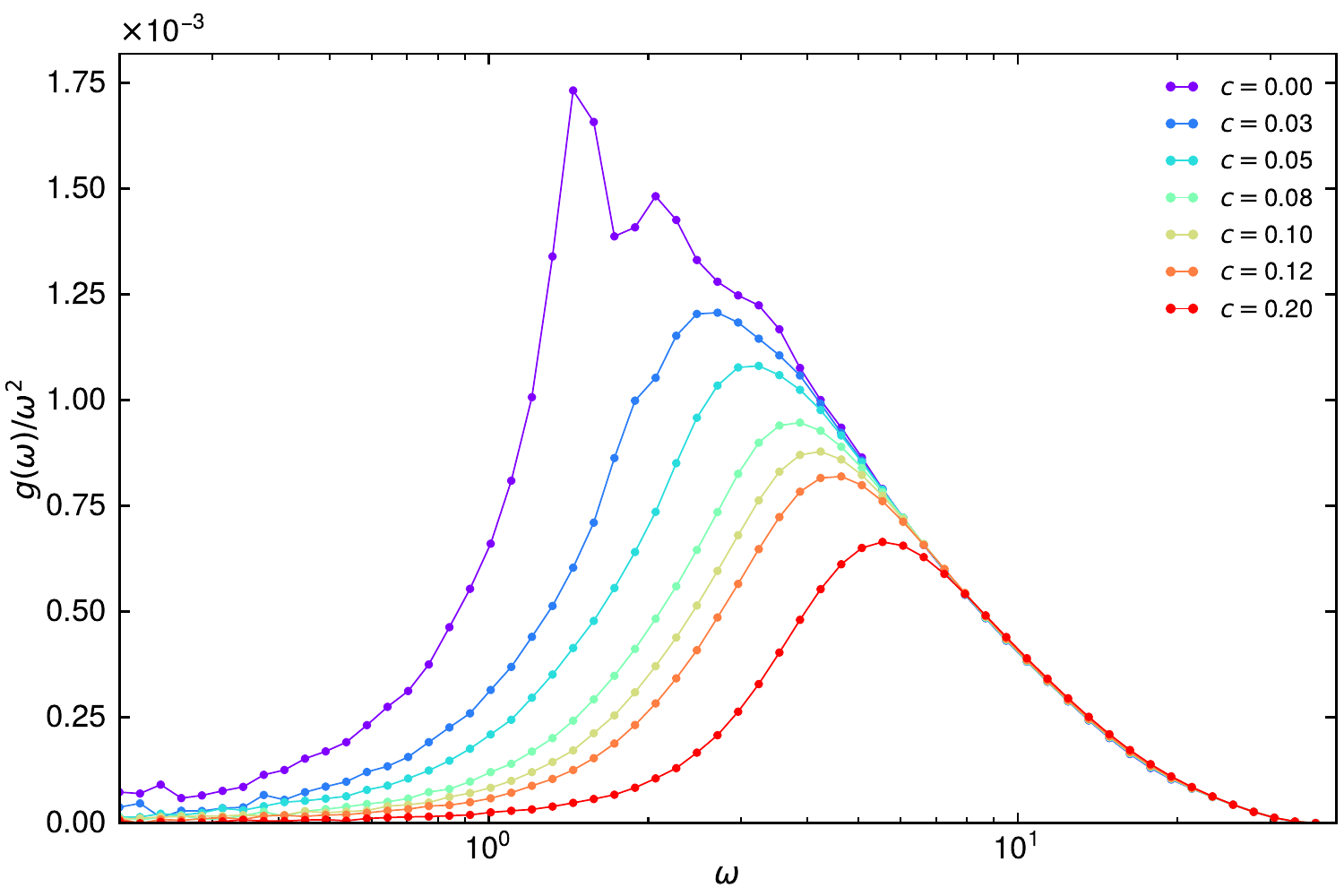}
\caption{
Reduced vibrational density of states $g(\omega)/\omega^2$ for several values of $c$.
Data exhibit BP for all cases of $c$.}
\label{fig:BP}
\end{figure}

We examine the participation ratio $p_k$ of each vibrational eigenmode $k$, which is defined as $p_k = 1/\pqty{N_\text{up}\sum_{i=1}^{N_\text{up}}\abs{\vb*{e}_k^i}^4}$ and quantifies the fraction of particles that participate in the vibration ($N_\text{up} p_k$ quantifies the number of participating particles)~\cite{Schober_1991,mazzacurati_1996}.
As in the extreme cases, $p_k = 1$ ($N_\text{up} p_k = N_\text{up}$) for an ideal mode in which all the unpinned particles vibrate equally, and $p_k = 1/N_\text{up} \ll 1$ ($N_\text{up} p_k = 1$) for an ideal mode involving only one particle.
Figure~\ref{fig:PR} presents data of $p_k$ versus $\omega_k$ for the configurations with (a) $c = 0.00$, (b) 0.05, (c) 0.10, and (d) 0.20.
Note that the thermodynamic glass transition occurs at approximately $c = 0.10$, and the system of $c = 0.20$ is located deep in the ideal glass phase.
From the figure, we clearly recognize that the localized modes survive and do \textit{not} vanish through the glass transition.
In addition, for all cases of $c$, localized modes always exist below the BP frequency $\omega_\text{BP}$ (see Fig.~\ref{fig:BP} for the BP and $\omega_\text{BP}$).

\begin{figure}
\centering
\includegraphics[width=\linewidth]{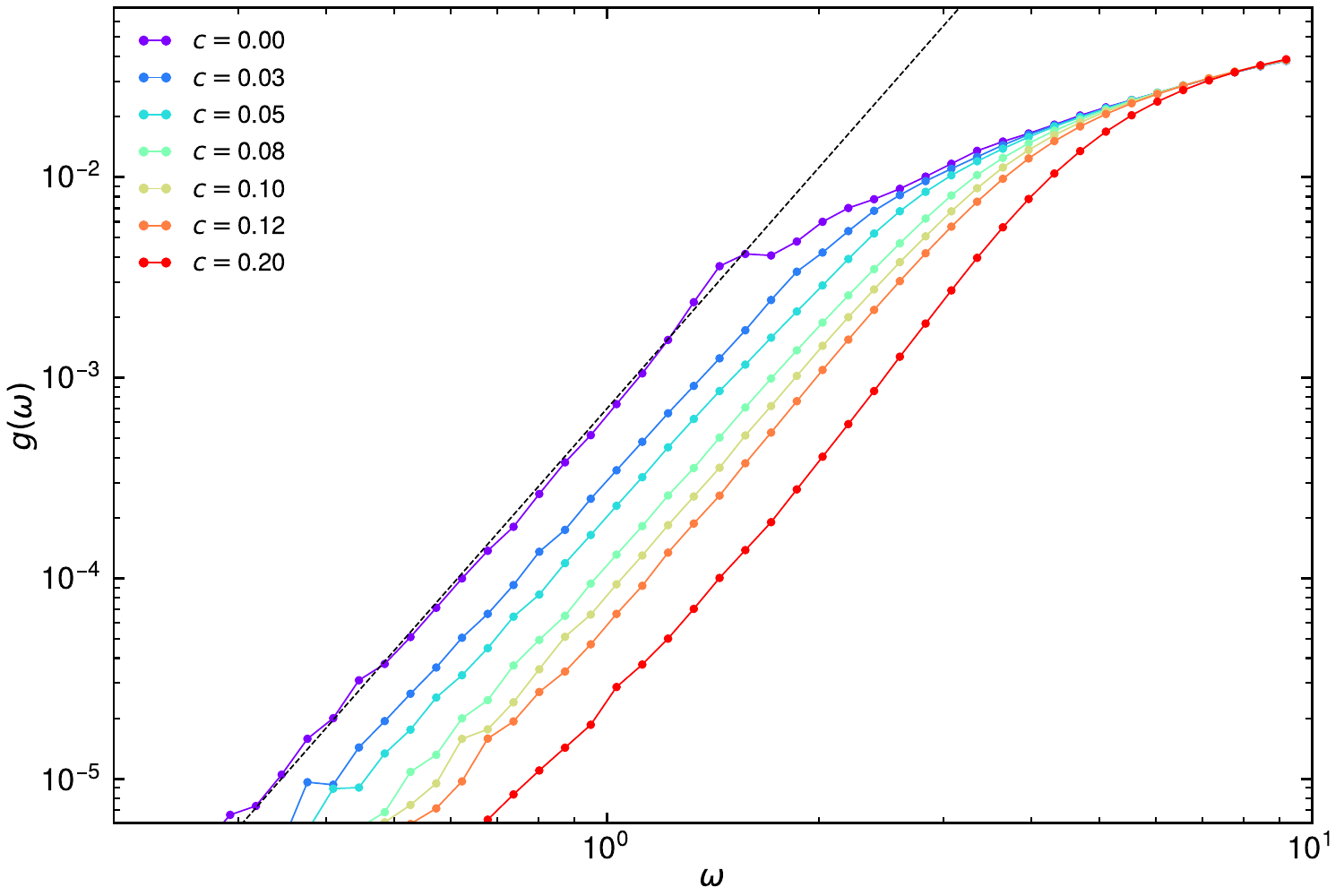}
\caption{
The vibrational density of states for several different values of $c$.
System is composed of $N = 4000$ particles.
Dashed line indicates scaling law of $g(\omega) \propto \omega^4$.}
\label{fig:VDOS}
\end{figure}

To quantitatively see the changes in the number of these localized modes with $c$, particularly across the glass transition, we measure the VDOS $g(\omega)$.
Figure~\ref{fig:VDOS} presents data of $g(\omega)$ for several different values of $c$ from the unpinned case of $c = 0.00$ to the equilibrium glass cases of $c > 0.10$.
(The number of configurations used for this analysis is given in the Supplemental Material.)
The figure clearly demonstrates the $\omega^4$ scaling law for all cases of $c$.
Here, we note that there are some finite-size effects found in the calculation of $g(\omega)$~\cite{Lerner_PRE_2020}. When $c = 0.00$, the VDOS follows $g(\omega) \propto \omega^{3.5}$ for $N = 1000$ (data shown in Supplemental Material) and $g(\omega) \propto \omega^4$ for $N = 4000$, as shown in Fig.~\ref{fig:VDOS}. However, for the equilibrium glasses of $c > 0.10$, we do not see such size effects.
It is worth emphasizing that even deep in the equilibrium glass state at $c = 0.20$, the VDOS follows $g(\omega) \propto \omega^4$.

\begin{figure}
\centering
\includegraphics[width=\linewidth]{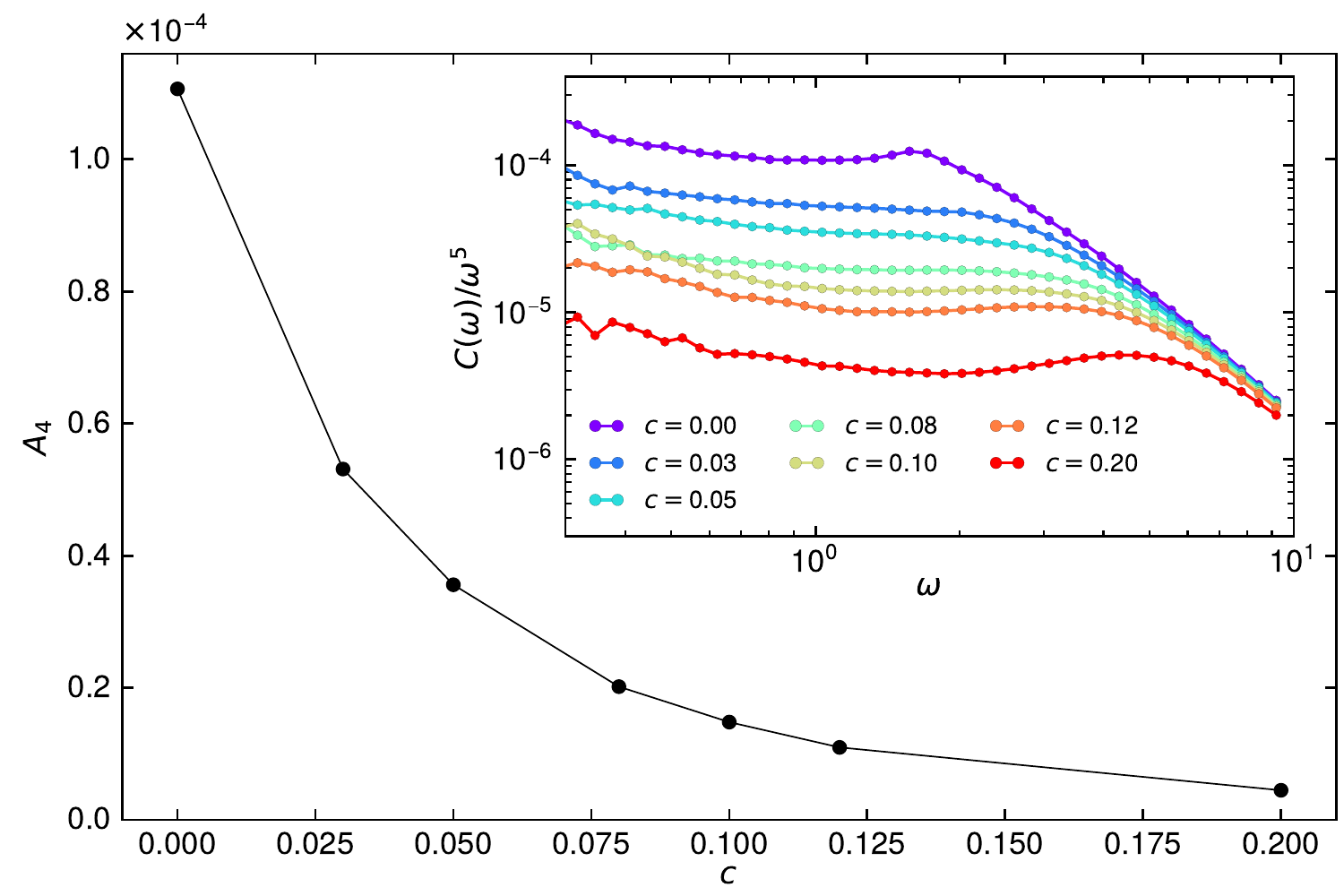}
\caption{
Coefficient $A_4$ in $\omega^4$ scaling law, $g(\omega) = A_4 \omega^4$.
$A_4$ is plotted as a function of $c$.
Inset shows reduced cumulative distribution function $C(\omega)/\omega^5$ for several values of $c$.}
\label{fig:A4}
\end{figure}

Since the VDOS always takes the form of $g(\omega) = A_4 \omega^4$, we measure how the coefficient $A_4$ depends on $c$.
To determine the value of $A_4$ precisely, we calculate the cumulative distribution function $C(\omega) = \int_0^\omega g(\omega^\prime) d\omega^\prime$ (see the inset of Fig.~\ref{fig:A4}).
We evaluate the height of the low-frequency plateau of $C(\omega)/\omega^5$ in the region of $0.7 < \omega < 1.5$ as $A_4$.
Figure~\ref{fig:A4} plots $A_4$ as a function of $c$.
From the figure, we see that $A_4$ decreases monotonically with increasing $c$.
Note that the dependence of $A_4$ on $c$ is always continuous without any signal of discontinuities, although the system crosses the thermodynamic glass transition to the ideal glass state.
Therefore, we conclude that the localized vibrations change continuously while maintaining the $\omega^4$ scaling law.
This behavior also indicates that the bottom of the potential energy landscape deforms smoothly through the thermodynamic glass transition.
In Ref.~\cite{Niblett2018}, it is revealed that the increase in the fraction of pinned particles corresponds to the change in the global landscape from a multiple-metabasin structure to a single-funnel structure.
Our result suggests that the local structure of the bottoms of basins is shared throughout this change in the landscape structure, reflecting the thermodynamic glass transition.

Next, we focus on the BP, the excess of the vibrational modes to the Debye prediction, which is recognized as the excess value in the reduced VDOS $g(\omega)/\omega^2$ and is ubiquitously observed in many glasses~\cite{Buchenau_1984,Yamamuro_1996,Mori_2020}.
Figure~\ref{fig:BP} presents data of the reduced VDOS and demonstrates that the BP always persists through the thermodynamic glass transition, as do the localized modes.
As shown in Fig.~\ref{fig:BP}, the peak height gradually decreases, and the peak frequency $\omega_\text{BP}$ increases as $c$ increases.
This behavior is consistent with observations at higher temperatures $T_p = 3T_c$~\cite{Angelani_2018}, but the present case crosses the glass transition.
An experimental study of polymeric glass former reported that the BP can disappear in the ideal glass states~\cite{Monnier_2021}.
However, our result suggests that BP can persist even in ideal glass states.

It has been understood that the localized modes originate from BP~\cite{Wyart_2006,Xu_2007}.
The vibrational modes in the BP regime decrease to the lower-frequency regime to become the localized modes due to the effects of the repulsive interactions between particles.
Thus, the BP and localized modes exist concurrently, and the localized modes appear below the BP frequency~\cite{Mizuno_2017}.
From Figs.~\ref{fig:PR} and~\ref{fig:BP}, we can confirm this feature is also true for the present equilibrium glasses.
Note that in particulate gels that are low-density amorphous solids, neither the BP nor the localized modes are observed~\cite{Mizuno_2021}.

\begin{figure}
\centering
\includegraphics[width=\linewidth]{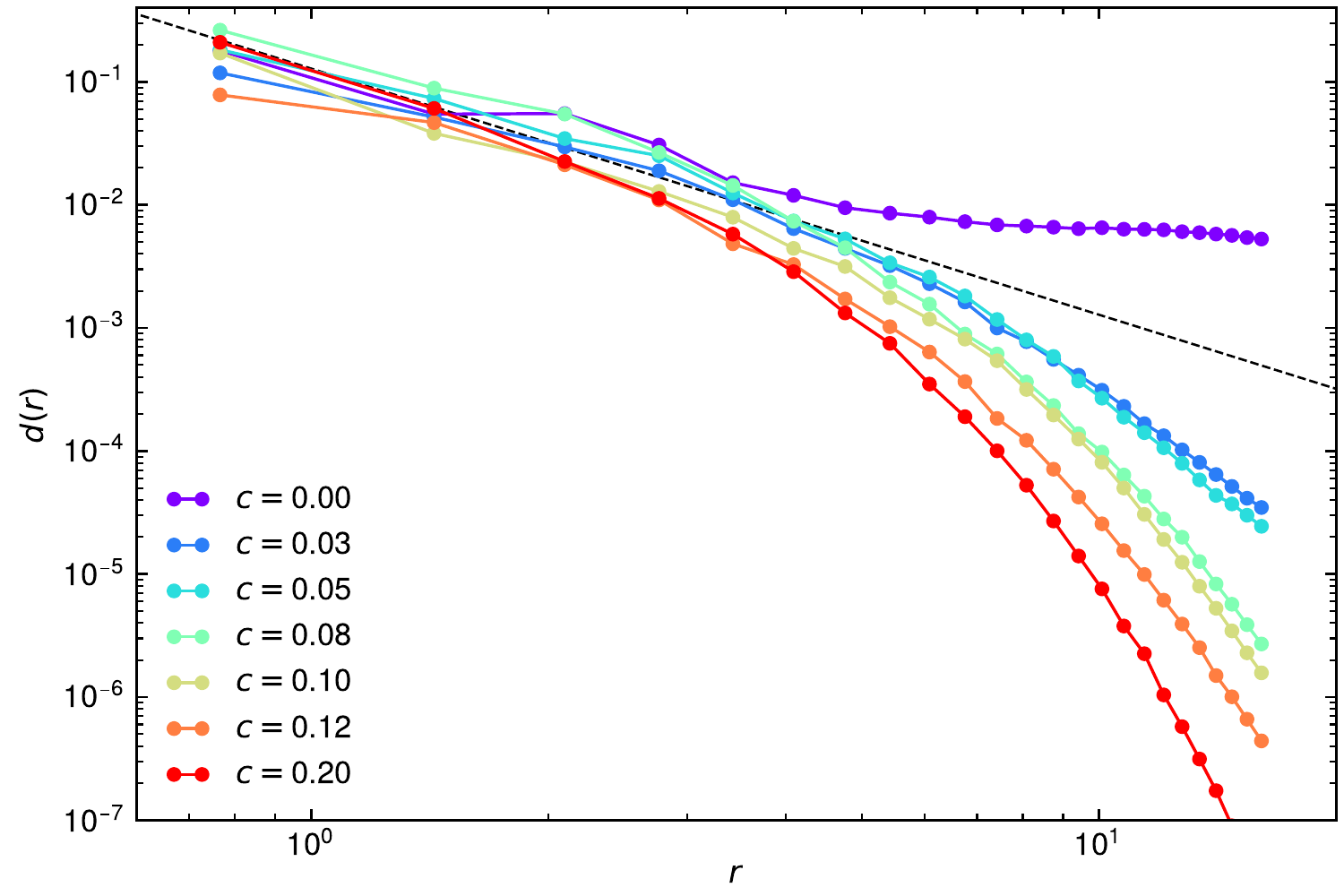}
\caption{
Decay profile $d(r)$ of lowest localized mode for several values of $c$.
System is composed of $N = 40000$ particles.
Note that when $c = 0.00$, we pick sixth lowest frequency mode with $p_k = 2.99 \times 10^{-4}$.
Dashed line indicates power-law behavior of $d(r) \propto r^{-2}$.}
\label{fig:decay}
\end{figure}

Finally, we probe the spatial structure of the localized modes.
We pick up the ``core'' particle that vibrates with the largest displacement $\abs{\vb*{e}_k^i}$ and then measure how displacements $\abs{\vb*{e}_k^i}$ of the other particles decay with distance $r$ from the core particle.
Therefore, we calculate $d(r) = \abs{\vb*{e}_k^i}/\argmax_i \abs{\vb*{e}_k^i}$ as a function of $r$.
(Note that we take the median of each contribution $\abs{\vb*{e}_k^i}$ from particles inside a shell with radius $r$.)
Figure~\ref{fig:decay} plots $d(r)$ as a function of $r$ for several different values of $c$.
It is well-known that for the unpinned system of $c = 0.00$, the localized modes hybridize with phonons, which leads to the power-law decay of $d(r) \propto r^{-2}$ (see the dashed line; note that this behavior is obscured by the finite-size effect in Fig.~\ref{fig:decay})~\cite{Lerner_2016}.
However, in the pinned systems of $c > 0$, $d(r)$ follows rather steep (possibly exponential) decay.
This is because phonons are suppressed in the pinned systems due to the break of translational invariance~\cite{Angelani_2018}.
(This point can be recognized in data of the participation ratio in Fig.~\ref{fig:PR}, i.e., we observe a ``tip'' at $c = 0.00$ that corresponds to the phonons, whereas the tip is totally absent at $c = 0.20$.)
Thus, we can confirm from the data of $d(r)$ that the pinned systems show ``bare'' or ``truly'' localized modes without hybridization with phonons.

In summary, we study ideal glasses using the random pinning method.
We find that the localized vibrational modes and the BP continuously evolve through the thermodynamic glass transition and survive even in ideal glass states.
Remarkably, the $\omega^4$ scaling law is robust through the glass transition.
Additionally, the localized modes always exist below the BP frequency $\omega_\text{BP}$, indicating that they originate from the BP modes.

In the present work, we used the random pinning method, which possesses two advantages.
First, it can shift the glass transition to a higher temperature region~\cite{Cammarota_2012,Kob_2013,Ozawa_2015} such that we can simulate the system experiencing the thermodynamic glass transition within a reasonable computational cost.
Second, it can suppress phonons in the system and solve hybridizations of localized vibrations with the phonons~\cite{Angelani_2018} such that we can analyze the bare localized vibrations.
This second point is important since hybridization with phonons can induce harmful consequences, e.g., finite-size effects in the calculation of the VDOS~\cite{Lerner_PRE_2020} (see also the most recent remark~\cite{Lerner_2022}).

In addition, thanks to the suppression of phonons, we can focus directly on effects from the bare localized modes in glasses.
Recent work on plasticity~\cite{Bhowmik2019Effect} indicated that the yielding of pinned glasses is governed by intensive events without brittle stress drops.
They also found that system-spanning rearrangement events are totally absent during plastic events.
These properties of yielding, which are caused by the bare localized modes, are markedly different from those in the unpinned glasses where the localized modes hybridize with phonons.
It will be interesting in the future to study the role of bare localized vibrations in other properties, e.g., the anharmonic contribution of entropy~\cite{Ozawa_PRL_2018} and phonon transport~\cite{Mizuno_2018}.
Additionally, to disentangle localized vibrations from phonons, another method was proposed to implement an artificial potential that acts as a high-pass filter~\cite{Wijtmans_2017}.

We emphasize that the $\omega^4$ scaling law is very robust, persisting in the ideal glass states.
This scaling persists whether the localized vibrations hybridize with the phonons (as in unpinned systems) or not (as in pinned systems).
We can also find $\omega^4$ scaling in many types of glasses~\cite{Richard_2020,Zapperi_2020,Shiraishi_2020}.
This robustness may occur because the $\omega^4$ scaling can be explained by simple phenomenological arguments~\cite{Gurarie_2003,Gurevich_2003} as well as the EMT framework~\cite{Shimada:2020ka,Shimada:2020vi,Shimada:2020vc}.

As a final remark, the present results provide useful insight into the material properties of equilibrium glasses.
Many past works studied experimental, nonequilibrium glasses and elucidated that the low-frequency vibrational modes play central roles in, e.g., responses under shear deformations~\cite{Tanguy_2010,Manning_2011,Oyama2021HB}, heterogeneous thermal relaxations~\cite{Widmer_Cooper_2008}, nonaffine responses of elasticity~\cite{Lemaitre_2006,Mizuno_Saitoh_Silbert_2016}, and two-level systems~\cite{Anderson_1972,Phillips_1987,Khomenko_2020}.
Since the low-frequency modes also survive in ideal glasses, we expect that equilibrium glasses share the properties of nonequilibrium glasses.

\begin{acknowledgments}
We thank Misaki Ozawa and Atsushi Ikeda for useful discussions.
This work is supported by JSPS KAKENHI (Grant Nos.~18H05225, 19H01812, 20H00128, 20H01868, 21J10021, 22K03543), JST SPRING (Grant No.~JPMJSP2108), and the Initiative on Promotion of Supercomputing for Young or Women Researchers, Information Technology Center, the University of Tokyo.
\end{acknowledgments}

\begin{thebibliography}{66}%
\makeatletter
\providecommand \@ifxundefined [1]{%
 \@ifx{#1\undefined}
}%
\providecommand \@ifnum [1]{%
 \ifnum #1\expandafter \@firstoftwo
 \else \expandafter \@secondoftwo
 \fi
}%
\providecommand \@ifx [1]{%
 \ifx #1\expandafter \@firstoftwo
 \else \expandafter \@secondoftwo
 \fi
}%
\providecommand \natexlab [1]{#1}%
\providecommand \enquote  [1]{``#1''}%
\providecommand \bibnamefont  [1]{#1}%
\providecommand \bibfnamefont [1]{#1}%
\providecommand \citenamefont [1]{#1}%
\providecommand \href@noop [0]{\@secondoftwo}%
\providecommand \href [0]{\begingroup \@sanitize@url \@href}%
\providecommand \@href[1]{\@@startlink{#1}\@@href}%
\providecommand \@@href[1]{\endgroup#1\@@endlink}%
\providecommand \@sanitize@url [0]{\catcode `\\12\catcode `\$12\catcode
  `\&12\catcode `\#12\catcode `\^12\catcode `\_12\catcode `\%12\relax}%
\providecommand \@@startlink[1]{}%
\providecommand \@@endlink[0]{}%
\providecommand \url  [0]{\begingroup\@sanitize@url \@url }%
\providecommand \@url [1]{\endgroup\@href {#1}{\urlprefix }}%
\providecommand \urlprefix  [0]{URL }%
\providecommand \Eprint [0]{\href }%
\providecommand \doibase [0]{https://doi.org/}%
\providecommand \selectlanguage [0]{\@gobble}%
\providecommand \bibinfo  [0]{\@secondoftwo}%
\providecommand \bibfield  [0]{\@secondoftwo}%
\providecommand \translation [1]{[#1]}%
\providecommand \BibitemOpen [0]{}%
\providecommand \bibitemStop [0]{}%
\providecommand \bibitemNoStop [0]{.\EOS\space}%
\providecommand \EOS [0]{\spacefactor3000\relax}%
\providecommand \BibitemShut  [1]{\csname bibitem#1\endcsname}%
\let\auto@bib@innerbib\@empty
\bibitem [{\citenamefont {DeGiuli}\ \emph {et~al.}(2014)\citenamefont
  {DeGiuli}, \citenamefont {Laversanne-Finot}, \citenamefont {D{\"u}ring},
  \citenamefont {Lerner},\ and\ \citenamefont {Wyart}}]{DeGiuli_2014}%
  \BibitemOpen
  \bibfield  {author} {\bibinfo {author} {\bibfnamefont {E.}~\bibnamefont
  {DeGiuli}}, \bibinfo {author} {\bibfnamefont {A.}~\bibnamefont
  {Laversanne-Finot}}, \bibinfo {author} {\bibfnamefont {G.}~\bibnamefont
  {D{\"u}ring}}, \bibinfo {author} {\bibfnamefont {E.}~\bibnamefont {Lerner}},\
  and\ \bibinfo {author} {\bibfnamefont {M.}~\bibnamefont {Wyart}},\ }\href
  {https://doi.org/10.1039/c4sm00561a} {\bibfield  {journal} {\bibinfo
  {journal} {Soft Matter}\ }\textbf {\bibinfo {volume} {10}},\ \bibinfo {pages}
  {5628} (\bibinfo {year} {2014})}\BibitemShut {NoStop}%
\bibitem [{\citenamefont {Franz}\ \emph {et~al.}(2015)\citenamefont {Franz},
  \citenamefont {Parisi}, \citenamefont {Urbani},\ and\ \citenamefont
  {Zamponi}}]{Franz_2015}%
  \BibitemOpen
  \bibfield  {author} {\bibinfo {author} {\bibfnamefont {S.}~\bibnamefont
  {Franz}}, \bibinfo {author} {\bibfnamefont {G.}~\bibnamefont {Parisi}},
  \bibinfo {author} {\bibfnamefont {P.}~\bibnamefont {Urbani}},\ and\ \bibinfo
  {author} {\bibfnamefont {F.}~\bibnamefont {Zamponi}},\ }\href
  {https://doi.org/10.1073/pnas.1511134112} {\bibfield  {journal} {\bibinfo
  {journal} {Proceedings of the National Academy of Sciences}\ }\textbf
  {\bibinfo {volume} {112}},\ \bibinfo {pages} {14539} (\bibinfo {year}
  {2015})}\BibitemShut {NoStop}%
\bibitem [{\citenamefont {Ashcroft}\ and\ \citenamefont
  {Mermin}(1976)}]{Ashcroft_Mermin}%
  \BibitemOpen
  \bibfield  {author} {\bibinfo {author} {\bibfnamefont {N.~W.}\ \bibnamefont
  {Ashcroft}}\ and\ \bibinfo {author} {\bibfnamefont {N.~D.}\ \bibnamefont
  {Mermin}},\ }\href@noop {} {\emph {\bibinfo {title} {{Solid State
  Physics}}}}\ (\bibinfo  {publisher} {Saunders College},\ \bibinfo {year}
  {1976})\BibitemShut {NoStop}%
\bibitem [{\citenamefont {Charbonneau}\ \emph {et~al.}(2016)\citenamefont
  {Charbonneau}, \citenamefont {Corwin}, \citenamefont {Parisi}, \citenamefont
  {Poncet},\ and\ \citenamefont {Zamponi}}]{Charbonneau_2016}%
  \BibitemOpen
  \bibfield  {author} {\bibinfo {author} {\bibfnamefont {P.}~\bibnamefont
  {Charbonneau}}, \bibinfo {author} {\bibfnamefont {E.~I.}\ \bibnamefont
  {Corwin}}, \bibinfo {author} {\bibfnamefont {G.}~\bibnamefont {Parisi}},
  \bibinfo {author} {\bibfnamefont {A.}~\bibnamefont {Poncet}},\ and\ \bibinfo
  {author} {\bibfnamefont {F.}~\bibnamefont {Zamponi}},\ }\href
  {https://doi.org/10.1103/physrevlett.117.045503} {\bibfield  {journal}
  {\bibinfo  {journal} {Physical Review Letters}\ }\textbf {\bibinfo {volume}
  {117}},\ \bibinfo {pages} {045503} (\bibinfo {year} {2016})}\BibitemShut
  {NoStop}%
\bibitem [{\citenamefont {Shimada}\ \emph
  {et~al.}(2020{\natexlab{a}})\citenamefont {Shimada}, \citenamefont {Mizuno},
  \citenamefont {Berthier},\ and\ \citenamefont {Ikeda}}]{Shimada_2020}%
  \BibitemOpen
  \bibfield  {author} {\bibinfo {author} {\bibfnamefont {M.}~\bibnamefont
  {Shimada}}, \bibinfo {author} {\bibfnamefont {H.}~\bibnamefont {Mizuno}},
  \bibinfo {author} {\bibfnamefont {L.}~\bibnamefont {Berthier}},\ and\
  \bibinfo {author} {\bibfnamefont {A.}~\bibnamefont {Ikeda}},\ }\href
  {https://doi.org/10.1103/physreve.101.052906} {\bibfield  {journal} {\bibinfo
   {journal} {Physical Review E}\ }\textbf {\bibinfo {volume} {101}},\ \bibinfo
  {pages} {052906} (\bibinfo {year} {2020}{\natexlab{a}})}\BibitemShut
  {NoStop}%
\bibitem [{\citenamefont {Lerner}\ \emph {et~al.}(2016)\citenamefont {Lerner},
  \citenamefont {D{\"u}ring},\ and\ \citenamefont {Bouchbinder}}]{Lerner_2016}%
  \BibitemOpen
  \bibfield  {author} {\bibinfo {author} {\bibfnamefont {E.}~\bibnamefont
  {Lerner}}, \bibinfo {author} {\bibfnamefont {G.}~\bibnamefont {D{\"u}ring}},\
  and\ \bibinfo {author} {\bibfnamefont {E.}~\bibnamefont {Bouchbinder}},\
  }\href {https://doi.org/10.1103/physrevlett.117.035501} {\bibfield  {journal}
  {\bibinfo  {journal} {Physical Review Letters}\ }\textbf {\bibinfo {volume}
  {117}},\ \bibinfo {pages} {035501} (\bibinfo {year} {2016})}\BibitemShut
  {NoStop}%
\bibitem [{\citenamefont {Mizuno}\ \emph {et~al.}(2017)\citenamefont {Mizuno},
  \citenamefont {Shiba},\ and\ \citenamefont {Ikeda}}]{Mizuno_2017}%
  \BibitemOpen
  \bibfield  {author} {\bibinfo {author} {\bibfnamefont {H.}~\bibnamefont
  {Mizuno}}, \bibinfo {author} {\bibfnamefont {H.}~\bibnamefont {Shiba}},\ and\
  \bibinfo {author} {\bibfnamefont {A.}~\bibnamefont {Ikeda}},\ }\href
  {https://doi.org/10.1073/pnas.1709015114} {\bibfield  {journal} {\bibinfo
  {journal} {Proceedings of the National Academy of Sciences}\ }\textbf
  {\bibinfo {volume} {114}},\ \bibinfo {pages} {E9767} (\bibinfo {year}
  {2017})}\BibitemShut {NoStop}%
\bibitem [{\citenamefont {Shimada}\ \emph
  {et~al.}(2018{\natexlab{a}})\citenamefont {Shimada}, \citenamefont {Mizuno},\
  and\ \citenamefont {Ikeda}}]{Shimada_LJ_2018}%
  \BibitemOpen
  \bibfield  {author} {\bibinfo {author} {\bibfnamefont {M.}~\bibnamefont
  {Shimada}}, \bibinfo {author} {\bibfnamefont {H.}~\bibnamefont {Mizuno}},\
  and\ \bibinfo {author} {\bibfnamefont {A.}~\bibnamefont {Ikeda}},\ }\href
  {https://doi.org/10.1103/physreve.97.022609} {\bibfield  {journal} {\bibinfo
  {journal} {Physical Review E}\ }\textbf {\bibinfo {volume} {97}},\ \bibinfo
  {pages} {022609} (\bibinfo {year} {2018}{\natexlab{a}})}\BibitemShut
  {NoStop}%
\bibitem [{\citenamefont {Shimada}\ \emph
  {et~al.}(2020{\natexlab{b}})\citenamefont {Shimada}, \citenamefont {Mizuno},\
  and\ \citenamefont {Ikeda}}]{Shimada:2020ka}%
  \BibitemOpen
  \bibfield  {author} {\bibinfo {author} {\bibfnamefont {M.}~\bibnamefont
  {Shimada}}, \bibinfo {author} {\bibfnamefont {H.}~\bibnamefont {Mizuno}},\
  and\ \bibinfo {author} {\bibfnamefont {A.}~\bibnamefont {Ikeda}},\ }\href
  {https://doi.org/10.1039/d0sm00376j} {\bibfield  {journal} {\bibinfo
  {journal} {Soft Matter}\ }\textbf {\bibinfo {volume} {16}},\ \bibinfo {pages}
  {7279} (\bibinfo {year} {2020}{\natexlab{b}})}\BibitemShut {NoStop}%
\bibitem [{\citenamefont {Shimada}\ \emph {et~al.}(2021)\citenamefont
  {Shimada}, \citenamefont {Mizuno},\ and\ \citenamefont
  {Ikeda}}]{Shimada:2020vi}%
  \BibitemOpen
  \bibfield  {author} {\bibinfo {author} {\bibfnamefont {M.}~\bibnamefont
  {Shimada}}, \bibinfo {author} {\bibfnamefont {H.}~\bibnamefont {Mizuno}},\
  and\ \bibinfo {author} {\bibfnamefont {A.}~\bibnamefont {Ikeda}},\ }\href
  {https://doi.org/10.1039/d0sm01583k} {\bibfield  {journal} {\bibinfo
  {journal} {Soft Matter}\ }\textbf {\bibinfo {volume} {17}},\ \bibinfo {pages}
  {346} (\bibinfo {year} {2021})}\BibitemShut {NoStop}%
\bibitem [{\citenamefont {Shimada}\ and\ \citenamefont
  {De~Giuli}(2022)}]{Shimada:2020vc}%
  \BibitemOpen
  \bibfield  {author} {\bibinfo {author} {\bibfnamefont {M.}~\bibnamefont
  {Shimada}}\ and\ \bibinfo {author} {\bibfnamefont {E.}~\bibnamefont
  {De~Giuli}},\ }\href {https://doi.org/10.21468/SciPostPhys.12.3.090}
  {\bibfield  {journal} {\bibinfo  {journal} {SciPost Physics}\ }\textbf
  {\bibinfo {volume} {12}},\ \bibinfo {pages} {090} (\bibinfo {year}
  {2022})}\BibitemShut {NoStop}%
\bibitem [{\citenamefont {Shimada}\ \emph
  {et~al.}(2018{\natexlab{b}})\citenamefont {Shimada}, \citenamefont {Mizuno},
  \citenamefont {Wyart},\ and\ \citenamefont {Ikeda}}]{Shimada_spatial_2018}%
  \BibitemOpen
  \bibfield  {author} {\bibinfo {author} {\bibfnamefont {M.}~\bibnamefont
  {Shimada}}, \bibinfo {author} {\bibfnamefont {H.}~\bibnamefont {Mizuno}},
  \bibinfo {author} {\bibfnamefont {M.}~\bibnamefont {Wyart}},\ and\ \bibinfo
  {author} {\bibfnamefont {A.}~\bibnamefont {Ikeda}},\ }\href
  {https://doi.org/10.1103/physreve.98.060901} {\bibfield  {journal} {\bibinfo
  {journal} {Physical Review E}\ }\textbf {\bibinfo {volume} {98}},\ \bibinfo
  {pages} {060901} (\bibinfo {year} {2018}{\natexlab{b}})}\BibitemShut
  {NoStop}%
\bibitem [{\citenamefont {Wang}\ \emph {et~al.}(2019)\citenamefont {Wang},
  \citenamefont {Ninarello}, \citenamefont {Guan}, \citenamefont {Berthier},
  \citenamefont {Szamel},\ and\ \citenamefont {Flenner}}]{Wang_low_freq_2019}%
  \BibitemOpen
  \bibfield  {author} {\bibinfo {author} {\bibfnamefont {L.}~\bibnamefont
  {Wang}}, \bibinfo {author} {\bibfnamefont {A.}~\bibnamefont {Ninarello}},
  \bibinfo {author} {\bibfnamefont {P.}~\bibnamefont {Guan}}, \bibinfo {author}
  {\bibfnamefont {L.}~\bibnamefont {Berthier}}, \bibinfo {author}
  {\bibfnamefont {G.}~\bibnamefont {Szamel}},\ and\ \bibinfo {author}
  {\bibfnamefont {E.}~\bibnamefont {Flenner}},\ }\href
  {https://doi.org/10.1038/s41467-018-07978-1} {\bibfield  {journal} {\bibinfo
  {journal} {Nature Communications}\ }\textbf {\bibinfo {volume} {10}},\
  \bibinfo {pages} {26} (\bibinfo {year} {2019})}\BibitemShut {NoStop}%
\bibitem [{\citenamefont {Rainone}\ \emph {et~al.}(2020)\citenamefont
  {Rainone}, \citenamefont {Bouchbinder},\ and\ \citenamefont
  {Lerner}}]{Rainone_PNAS_2020}%
  \BibitemOpen
  \bibfield  {author} {\bibinfo {author} {\bibfnamefont {C.}~\bibnamefont
  {Rainone}}, \bibinfo {author} {\bibfnamefont {E.}~\bibnamefont
  {Bouchbinder}},\ and\ \bibinfo {author} {\bibfnamefont {E.}~\bibnamefont
  {Lerner}},\ }\href {https://doi.org/10.1073/pnas.1919958117} {\bibfield
  {journal} {\bibinfo  {journal} {Proceedings of the National Academy of
  Sciences}\ }\textbf {\bibinfo {volume} {117}},\ \bibinfo {pages} {5228}
  (\bibinfo {year} {2020})}\BibitemShut {NoStop}%
\bibitem [{\citenamefont {Ji}\ \emph {et~al.}(2020)\citenamefont {Ji},
  \citenamefont {de~Geus}, \citenamefont {Popovi{\'c}}, \citenamefont
  {Agoritsas},\ and\ \citenamefont {Wyart}}]{Ji_2020}%
  \BibitemOpen
  \bibfield  {author} {\bibinfo {author} {\bibfnamefont {W.}~\bibnamefont
  {Ji}}, \bibinfo {author} {\bibfnamefont {T.~W.~J.}\ \bibnamefont {de~Geus}},
  \bibinfo {author} {\bibfnamefont {M.}~\bibnamefont {Popovi{\'c}}}, \bibinfo
  {author} {\bibfnamefont {E.}~\bibnamefont {Agoritsas}},\ and\ \bibinfo
  {author} {\bibfnamefont {M.}~\bibnamefont {Wyart}},\ }\href
  {https://doi.org/10.1103/physreve.102.062110} {\bibfield  {journal} {\bibinfo
   {journal} {Physical Review E}\ }\textbf {\bibinfo {volume} {102}},\ \bibinfo
  {pages} {062110} (\bibinfo {year} {2020})}\BibitemShut {NoStop}%
\bibitem [{\citenamefont {Ninarello}\ \emph {et~al.}(2017)\citenamefont
  {Ninarello}, \citenamefont {Berthier},\ and\ \citenamefont
  {Coslovich}}]{Ninarello_2017}%
  \BibitemOpen
  \bibfield  {author} {\bibinfo {author} {\bibfnamefont {A.}~\bibnamefont
  {Ninarello}}, \bibinfo {author} {\bibfnamefont {L.}~\bibnamefont
  {Berthier}},\ and\ \bibinfo {author} {\bibfnamefont {D.}~\bibnamefont
  {Coslovich}},\ }\href {https://doi.org/10.1103/physrevx.7.021039} {\bibfield
  {journal} {\bibinfo  {journal} {Physical Review X}\ }\textbf {\bibinfo
  {volume} {7}},\ \bibinfo {pages} {021039} (\bibinfo {year}
  {2017})}\BibitemShut {NoStop}%
\bibitem [{\citenamefont {Hukushima}\ and\ \citenamefont
  {Nemoto}(1996)}]{Hukushima_1996}%
  \BibitemOpen
  \bibfield  {author} {\bibinfo {author} {\bibfnamefont {K.}~\bibnamefont
  {Hukushima}}\ and\ \bibinfo {author} {\bibfnamefont {K.}~\bibnamefont
  {Nemoto}},\ }\href {https://doi.org/10.1143/JPSJ.65.1604} {\bibfield
  {journal} {\bibinfo  {journal} {Journal of the Physical Society of Japan}\
  }\textbf {\bibinfo {volume} {65}},\ \bibinfo {pages} {1604} (\bibinfo {year}
  {1996})}\BibitemShut {NoStop}%
\bibitem [{\citenamefont {Yamamoto}\ and\ \citenamefont
  {Kob}(2000)}]{Yamamoto_2000}%
  \BibitemOpen
  \bibfield  {author} {\bibinfo {author} {\bibfnamefont {R.}~\bibnamefont
  {Yamamoto}}\ and\ \bibinfo {author} {\bibfnamefont {W.}~\bibnamefont {Kob}},\
  }\href {https://doi.org/10.1103/PhysRevE.61.5473} {\bibfield  {journal}
  {\bibinfo  {journal} {Physical Review E}\ }\textbf {\bibinfo {volume} {61}},\
  \bibinfo {pages} {5473} (\bibinfo {year} {2000})}\BibitemShut {NoStop}%
\bibitem [{\citenamefont {De~Michele}\ and\ \citenamefont
  {Sciortino}(2002)}]{DeMichele2002}%
  \BibitemOpen
  \bibfield  {author} {\bibinfo {author} {\bibfnamefont {C.}~\bibnamefont
  {De~Michele}}\ and\ \bibinfo {author} {\bibfnamefont {F.}~\bibnamefont
  {Sciortino}},\ }\href {https://doi.org/10.1103/physreve.65.051202} {\bibfield
   {journal} {\bibinfo  {journal} {Physical Review E}\ }\textbf {\bibinfo
  {volume} {65}},\ \bibinfo {pages} {051202} (\bibinfo {year}
  {2002})}\BibitemShut {NoStop}%
\bibitem [{\citenamefont {Coslovich}\ \emph {et~al.}(2018)\citenamefont
  {Coslovich}, \citenamefont {Ozawa},\ and\ \citenamefont
  {Kob}}]{Coslovich_EPJE_2018}%
  \BibitemOpen
  \bibfield  {author} {\bibinfo {author} {\bibfnamefont {D.}~\bibnamefont
  {Coslovich}}, \bibinfo {author} {\bibfnamefont {M.}~\bibnamefont {Ozawa}},\
  and\ \bibinfo {author} {\bibfnamefont {W.}~\bibnamefont {Kob}},\ }\href
  {https://doi.org/10.1140/epje/i2018-11671-2} {\bibfield  {journal} {\bibinfo
  {journal} {The European Physical Journal E}\ }\textbf {\bibinfo {volume}
  {41}},\ \bibinfo {pages} {62} (\bibinfo {year} {2018})}\BibitemShut {NoStop}%
\bibitem [{\citenamefont {Grigera}\ and\ \citenamefont
  {Parisi}(2001)}]{Grigera_2001}%
  \BibitemOpen
  \bibfield  {author} {\bibinfo {author} {\bibfnamefont {T.~S.}\ \bibnamefont
  {Grigera}}\ and\ \bibinfo {author} {\bibfnamefont {G.}~\bibnamefont
  {Parisi}},\ }\href {https://doi.org/10.1103/physreve.63.045102} {\bibfield
  {journal} {\bibinfo  {journal} {Physical Review E}\ }\textbf {\bibinfo
  {volume} {63}},\ \bibinfo {pages} {045102} (\bibinfo {year}
  {2001})}\BibitemShut {NoStop}%
\bibitem [{\citenamefont {Guti{\'e}rrez}\ \emph {et~al.}(2015)\citenamefont
  {Guti{\'e}rrez}, \citenamefont {Karmakar}, \citenamefont {Pollack},\ and\
  \citenamefont {Procaccia}}]{Gutierrez_2015}%
  \BibitemOpen
  \bibfield  {author} {\bibinfo {author} {\bibfnamefont {R.}~\bibnamefont
  {Guti{\'e}rrez}}, \bibinfo {author} {\bibfnamefont {S.}~\bibnamefont
  {Karmakar}}, \bibinfo {author} {\bibfnamefont {Y.~G.}\ \bibnamefont
  {Pollack}},\ and\ \bibinfo {author} {\bibfnamefont {I.}~\bibnamefont
  {Procaccia}},\ }\href {https://doi.org/10.1209/0295-5075/111/56009}
  {\bibfield  {journal} {\bibinfo  {journal} {Europhysics Letters}\ }\textbf
  {\bibinfo {volume} {111}},\ \bibinfo {pages} {56009} (\bibinfo {year}
  {2015})}\BibitemShut {NoStop}%
\bibitem [{\citenamefont {Cammarota}\ and\ \citenamefont
  {Biroli}(2012)}]{Cammarota_2012}%
  \BibitemOpen
  \bibfield  {author} {\bibinfo {author} {\bibfnamefont {C.}~\bibnamefont
  {Cammarota}}\ and\ \bibinfo {author} {\bibfnamefont {G.}~\bibnamefont
  {Biroli}},\ }\href {https://doi.org/10.1073/pnas.1111582109} {\bibfield
  {journal} {\bibinfo  {journal} {Proceedings of the National Academy of
  Sciences}\ }\textbf {\bibinfo {volume} {109}},\ \bibinfo {pages} {8850}
  (\bibinfo {year} {2012})}\BibitemShut {NoStop}%
\bibitem [{\citenamefont {Ozawa}\ \emph {et~al.}(2015)\citenamefont {Ozawa},
  \citenamefont {Kob}, \citenamefont {Ikeda},\ and\ \citenamefont
  {Miyazaki}}]{Ozawa_2015}%
  \BibitemOpen
  \bibfield  {author} {\bibinfo {author} {\bibfnamefont {M.}~\bibnamefont
  {Ozawa}}, \bibinfo {author} {\bibfnamefont {W.}~\bibnamefont {Kob}}, \bibinfo
  {author} {\bibfnamefont {A.}~\bibnamefont {Ikeda}},\ and\ \bibinfo {author}
  {\bibfnamefont {K.}~\bibnamefont {Miyazaki}},\ }\href
  {https://doi.org/10.1073/pnas.1500730112} {\bibfield  {journal} {\bibinfo
  {journal} {Proceedings of the National Academy of Sciences}\ }\textbf
  {\bibinfo {volume} {112}},\ \bibinfo {pages} {6914} (\bibinfo {year}
  {2015})}\BibitemShut {NoStop}%
\bibitem [{\citenamefont {Kob}\ and\ \citenamefont
  {Berthier}(2013)}]{Kob_2013}%
  \BibitemOpen
  \bibfield  {author} {\bibinfo {author} {\bibfnamefont {W.}~\bibnamefont
  {Kob}}\ and\ \bibinfo {author} {\bibfnamefont {L.}~\bibnamefont {Berthier}},\
  }\href {https://doi.org/10.1103/physrevlett.110.245702} {\bibfield  {journal}
  {\bibinfo  {journal} {Physical Review Letters}\ }\textbf {\bibinfo {volume}
  {110}},\ \bibinfo {pages} {245702} (\bibinfo {year} {2013})}\BibitemShut
  {NoStop}%
\bibitem [{\citenamefont {Kim}(2003)}]{Kim_2003}%
  \BibitemOpen
  \bibfield  {author} {\bibinfo {author} {\bibfnamefont {K.}~\bibnamefont
  {Kim}},\ }\href {https://doi.org/10.1209/epl/i2003-00303-0} {\bibfield
  {journal} {\bibinfo  {journal} {Europhysics Letters}\ }\textbf {\bibinfo
  {volume} {61}},\ \bibinfo {pages} {790} (\bibinfo {year} {2003})}\BibitemShut
  {NoStop}%
\bibitem [{\citenamefont {Kim}\ \emph {et~al.}(2011)\citenamefont {Kim},
  \citenamefont {Miyazaki},\ and\ \citenamefont {Saito}}]{Kim_2011}%
  \BibitemOpen
  \bibfield  {author} {\bibinfo {author} {\bibfnamefont {K.}~\bibnamefont
  {Kim}}, \bibinfo {author} {\bibfnamefont {K.}~\bibnamefont {Miyazaki}},\ and\
  \bibinfo {author} {\bibfnamefont {S.}~\bibnamefont {Saito}},\ }\href
  {https://doi.org/10.1088/0953-8984/23/23/234123} {\bibfield  {journal}
  {\bibinfo  {journal} {Journal of Physics: Condensed Matter}\ }\textbf
  {\bibinfo {volume} {23}},\ \bibinfo {pages} {234123} (\bibinfo {year}
  {2011})}\BibitemShut {NoStop}%
\bibitem [{\citenamefont {Jack}\ and\ \citenamefont
  {Fullerton}(2013)}]{Jack2013}%
  \BibitemOpen
  \bibfield  {author} {\bibinfo {author} {\bibfnamefont {R.~L.}\ \bibnamefont
  {Jack}}\ and\ \bibinfo {author} {\bibfnamefont {C.~J.}\ \bibnamefont
  {Fullerton}},\ }\href {https://doi.org/10.1103/PhysRevE.88.042304} {\bibfield
   {journal} {\bibinfo  {journal} {Physical Review E}\ }\textbf {\bibinfo
  {volume} {88}},\ \bibinfo {pages} {042304} (\bibinfo {year}
  {2013})}\BibitemShut {NoStop}%
\bibitem [{\citenamefont {Kob}\ and\ \citenamefont
  {Coslovich}(2014)}]{Kob_2014}%
  \BibitemOpen
  \bibfield  {author} {\bibinfo {author} {\bibfnamefont {W.}~\bibnamefont
  {Kob}}\ and\ \bibinfo {author} {\bibfnamefont {D.}~\bibnamefont
  {Coslovich}},\ }\href {https://doi.org/10.1103/physreve.90.052305} {\bibfield
   {journal} {\bibinfo  {journal} {Physical Review E}\ }\textbf {\bibinfo
  {volume} {90}},\ \bibinfo {pages} {052305} (\bibinfo {year}
  {2014})}\BibitemShut {NoStop}%
\bibitem [{\citenamefont {Chakrabarty}\ \emph {et~al.}(2015)\citenamefont
  {Chakrabarty}, \citenamefont {Karmakar},\ and\ \citenamefont
  {Dasgupta}}]{Chakrabarty_2015}%
  \BibitemOpen
  \bibfield  {author} {\bibinfo {author} {\bibfnamefont {S.}~\bibnamefont
  {Chakrabarty}}, \bibinfo {author} {\bibfnamefont {S.}~\bibnamefont
  {Karmakar}},\ and\ \bibinfo {author} {\bibfnamefont {C.}~\bibnamefont
  {Dasgupta}},\ }\href {https://doi.org/10.1038/srep12577} {\bibfield
  {journal} {\bibinfo  {journal} {Scientific Reports}\ }\textbf {\bibinfo
  {volume} {5}},\ \bibinfo {pages} {12577} (\bibinfo {year}
  {2015})}\BibitemShut {NoStop}%
\bibitem [{\citenamefont {Chakrabarty}\ \emph {et~al.}(2016)\citenamefont
  {Chakrabarty}, \citenamefont {Das}, \citenamefont {Karmakar},\ and\
  \citenamefont {Dasgupta}}]{Chakrabarty2016}%
  \BibitemOpen
  \bibfield  {author} {\bibinfo {author} {\bibfnamefont {S.}~\bibnamefont
  {Chakrabarty}}, \bibinfo {author} {\bibfnamefont {R.}~\bibnamefont {Das}},
  \bibinfo {author} {\bibfnamefont {S.}~\bibnamefont {Karmakar}},\ and\
  \bibinfo {author} {\bibfnamefont {C.}~\bibnamefont {Dasgupta}},\ }\href
  {https://doi.org/10.1063/1.4958632} {\bibfield  {journal} {\bibinfo
  {journal} {The Journal of Chemical Physics}\ }\textbf {\bibinfo {volume}
  {145}},\ \bibinfo {pages} {034507} (\bibinfo {year} {2016})}\BibitemShut
  {NoStop}%
\bibitem [{\citenamefont {Ozawa}\ \emph {et~al.}(2018)\citenamefont {Ozawa},
  \citenamefont {Ikeda}, \citenamefont {Miyazaki},\ and\ \citenamefont
  {Kob}}]{Ozawa_PRL_2018}%
  \BibitemOpen
  \bibfield  {author} {\bibinfo {author} {\bibfnamefont {M.}~\bibnamefont
  {Ozawa}}, \bibinfo {author} {\bibfnamefont {A.}~\bibnamefont {Ikeda}},
  \bibinfo {author} {\bibfnamefont {K.}~\bibnamefont {Miyazaki}},\ and\
  \bibinfo {author} {\bibfnamefont {W.}~\bibnamefont {Kob}},\ }\href
  {https://doi.org/10.1103/physrevlett.121.205501} {\bibfield  {journal}
  {\bibinfo  {journal} {Physical Review Letters}\ }\textbf {\bibinfo {volume}
  {121}},\ \bibinfo {pages} {205501} (\bibinfo {year} {2018})}\BibitemShut
  {NoStop}%
\bibitem [{\citenamefont {Gokhale}\ \emph {et~al.}(2014)\citenamefont
  {Gokhale}, \citenamefont {Hima~Nagamanasa}, \citenamefont {Ganapathy},\ and\
  \citenamefont {Sood}}]{Gokhale_2014}%
  \BibitemOpen
  \bibfield  {author} {\bibinfo {author} {\bibfnamefont {S.}~\bibnamefont
  {Gokhale}}, \bibinfo {author} {\bibfnamefont {K.}~\bibnamefont
  {Hima~Nagamanasa}}, \bibinfo {author} {\bibfnamefont {R.}~\bibnamefont
  {Ganapathy}},\ and\ \bibinfo {author} {\bibfnamefont {A.~K.}\ \bibnamefont
  {Sood}},\ }\href {https://doi.org/10.1038/ncomms5685} {\bibfield  {journal}
  {\bibinfo  {journal} {Nature Communications}\ }\textbf {\bibinfo {volume}
  {5}},\ \bibinfo {pages} {4685} (\bibinfo {year} {2014})}\BibitemShut
  {NoStop}%
\bibitem [{\citenamefont {Angelani}\ \emph {et~al.}(2018)\citenamefont
  {Angelani}, \citenamefont {Paoluzzi}, \citenamefont {Parisi},\ and\
  \citenamefont {Ruocco}}]{Angelani_2018}%
  \BibitemOpen
  \bibfield  {author} {\bibinfo {author} {\bibfnamefont {L.}~\bibnamefont
  {Angelani}}, \bibinfo {author} {\bibfnamefont {M.}~\bibnamefont {Paoluzzi}},
  \bibinfo {author} {\bibfnamefont {G.}~\bibnamefont {Parisi}},\ and\ \bibinfo
  {author} {\bibfnamefont {G.}~\bibnamefont {Ruocco}},\ }\href
  {https://doi.org/10.1073/pnas.1805024115} {\bibfield  {journal} {\bibinfo
  {journal} {Proceedings of the National Academy of Sciences}\ }\textbf
  {\bibinfo {volume} {115}},\ \bibinfo {pages} {8700} (\bibinfo {year}
  {2018})}\BibitemShut {NoStop}%
\bibitem [{\citenamefont {Kob}\ and\ \citenamefont
  {Andersen}(1995)}]{Kob_Andersen_I_1995}%
  \BibitemOpen
  \bibfield  {author} {\bibinfo {author} {\bibfnamefont {W.}~\bibnamefont
  {Kob}}\ and\ \bibinfo {author} {\bibfnamefont {H.~C.}\ \bibnamefont
  {Andersen}},\ }\href {https://doi.org/10.1103/PhysRevE.51.4626} {\bibfield
  {journal} {\bibinfo  {journal} {Physical Review E}\ }\textbf {\bibinfo
  {volume} {51}},\ \bibinfo {pages} {4626} (\bibinfo {year}
  {1995})}\BibitemShut {NoStop}%
\bibitem [{\citenamefont {Kob}\ and\ \citenamefont
  {Andersen}(1994)}]{Kob_Andersen_PRL_1994}%
  \BibitemOpen
  \bibfield  {author} {\bibinfo {author} {\bibfnamefont {W.}~\bibnamefont
  {Kob}}\ and\ \bibinfo {author} {\bibfnamefont {H.~C.}\ \bibnamefont
  {Andersen}},\ }\href {https://doi.org/10.1103/PhysRevLett.73.1376} {\bibfield
   {journal} {\bibinfo  {journal} {Physical Review Letters}\ }\textbf {\bibinfo
  {volume} {73}},\ \bibinfo {pages} {1376} (\bibinfo {year}
  {1994})}\BibitemShut {NoStop}%
\bibitem [{\citenamefont {Gu{\'e}nol{\'e}}\ \emph {et~al.}(2020)\citenamefont
  {Gu{\'e}nol{\'e}}, \citenamefont {N{\"o}hring}, \citenamefont {Vaid},
  \citenamefont {Houll{\'e}}, \citenamefont {Xie}, \citenamefont {Prakash},\
  and\ \citenamefont {Bitzek}}]{Guenole_2020}%
  \BibitemOpen
  \bibfield  {author} {\bibinfo {author} {\bibfnamefont {J.}~\bibnamefont
  {Gu{\'e}nol{\'e}}}, \bibinfo {author} {\bibfnamefont {W.~G.}\ \bibnamefont
  {N{\"o}hring}}, \bibinfo {author} {\bibfnamefont {A.}~\bibnamefont {Vaid}},
  \bibinfo {author} {\bibfnamefont {F.}~\bibnamefont {Houll{\'e}}}, \bibinfo
  {author} {\bibfnamefont {Z.}~\bibnamefont {Xie}}, \bibinfo {author}
  {\bibfnamefont {A.}~\bibnamefont {Prakash}},\ and\ \bibinfo {author}
  {\bibfnamefont {E.}~\bibnamefont {Bitzek}},\ }\href
  {https://doi.org/10.1016/j.commatsci.2020.109584} {\bibfield  {journal}
  {\bibinfo  {journal} {Computational Materials Science}\ }\textbf {\bibinfo
  {volume} {175}},\ \bibinfo {pages} {109584} (\bibinfo {year}
  {2020})}\BibitemShut {NoStop}%
\bibitem [{\citenamefont {Schober}\ and\ \citenamefont
  {Laird}(1991)}]{Schober_1991}%
  \BibitemOpen
  \bibfield  {author} {\bibinfo {author} {\bibfnamefont {H.~R.}\ \bibnamefont
  {Schober}}\ and\ \bibinfo {author} {\bibfnamefont {B.~B.}\ \bibnamefont
  {Laird}},\ }\href {https://doi.org/10.1103/PhysRevB.44.6746} {\bibfield
  {journal} {\bibinfo  {journal} {Physical Review B}\ }\textbf {\bibinfo
  {volume} {44}},\ \bibinfo {pages} {6746} (\bibinfo {year}
  {1991})}\BibitemShut {NoStop}%
\bibitem [{\citenamefont {Mazzacurati}\ \emph {et~al.}(1996)\citenamefont
  {Mazzacurati}, \citenamefont {Ruocco},\ and\ \citenamefont
  {Sampoli}}]{mazzacurati_1996}%
  \BibitemOpen
  \bibfield  {author} {\bibinfo {author} {\bibfnamefont {V.}~\bibnamefont
  {Mazzacurati}}, \bibinfo {author} {\bibfnamefont {G.}~\bibnamefont
  {Ruocco}},\ and\ \bibinfo {author} {\bibfnamefont {M.}~\bibnamefont
  {Sampoli}},\ }\href {https://doi.org/10.1209/epl/i1996-00515-8} {\bibfield
  {journal} {\bibinfo  {journal} {Europhysics Letters}\ }\textbf {\bibinfo
  {volume} {34}},\ \bibinfo {pages} {681} (\bibinfo {year} {1996})}\BibitemShut
  {NoStop}%
\bibitem [{\citenamefont {Lerner}(2020)}]{Lerner_PRE_2020}%
  \BibitemOpen
  \bibfield  {author} {\bibinfo {author} {\bibfnamefont {E.}~\bibnamefont
  {Lerner}},\ }\href {https://doi.org/10.1103/physreve.101.032120} {\bibfield
  {journal} {\bibinfo  {journal} {Physical Review E}\ }\textbf {\bibinfo
  {volume} {101}},\ \bibinfo {pages} {032120} (\bibinfo {year}
  {2020})}\BibitemShut {NoStop}%
\bibitem [{\citenamefont {Niblett}\ \emph {et~al.}(2018)\citenamefont
  {Niblett}, \citenamefont {de~Souza}, \citenamefont {Jack},\ and\
  \citenamefont {Wales}}]{Niblett2018}%
  \BibitemOpen
  \bibfield  {author} {\bibinfo {author} {\bibfnamefont {S.~P.}\ \bibnamefont
  {Niblett}}, \bibinfo {author} {\bibfnamefont {V.~K.}\ \bibnamefont
  {de~Souza}}, \bibinfo {author} {\bibfnamefont {R.~L.}\ \bibnamefont {Jack}},\
  and\ \bibinfo {author} {\bibfnamefont {D.~J.}\ \bibnamefont {Wales}},\ }\href
  {https://doi.org/10.1063/1.5042140} {\bibfield  {journal} {\bibinfo
  {journal} {The Journal of Chemical Physics}\ }\textbf {\bibinfo {volume}
  {149}},\ \bibinfo {pages} {114503} (\bibinfo {year} {2018})}\BibitemShut
  {NoStop}%
\bibitem [{\citenamefont {Buchenau}\ \emph {et~al.}(1984)\citenamefont
  {Buchenau}, \citenamefont {N{\"u}cker},\ and\ \citenamefont
  {Dianoux}}]{Buchenau_1984}%
  \BibitemOpen
  \bibfield  {author} {\bibinfo {author} {\bibfnamefont {U.}~\bibnamefont
  {Buchenau}}, \bibinfo {author} {\bibfnamefont {N.}~\bibnamefont
  {N{\"u}cker}},\ and\ \bibinfo {author} {\bibfnamefont {A.~J.}\ \bibnamefont
  {Dianoux}},\ }\href {https://doi.org/10.1103/PhysRevLett.53.2316} {\bibfield
  {journal} {\bibinfo  {journal} {Physical Review Letters}\ }\textbf {\bibinfo
  {volume} {53}},\ \bibinfo {pages} {2316} (\bibinfo {year}
  {1984})}\BibitemShut {NoStop}%
\bibitem [{\citenamefont {Yamamuro}\ \emph {et~al.}(1996)\citenamefont
  {Yamamuro}, \citenamefont {Matsuo}, \citenamefont {Takeda}, \citenamefont
  {Kanaya}, \citenamefont {Kawaguchi},\ and\ \citenamefont
  {Kaji}}]{Yamamuro_1996}%
  \BibitemOpen
  \bibfield  {author} {\bibinfo {author} {\bibfnamefont {O.}~\bibnamefont
  {Yamamuro}}, \bibinfo {author} {\bibfnamefont {T.}~\bibnamefont {Matsuo}},
  \bibinfo {author} {\bibfnamefont {K.}~\bibnamefont {Takeda}}, \bibinfo
  {author} {\bibfnamefont {T.}~\bibnamefont {Kanaya}}, \bibinfo {author}
  {\bibfnamefont {T.}~\bibnamefont {Kawaguchi}},\ and\ \bibinfo {author}
  {\bibfnamefont {K.}~\bibnamefont {Kaji}},\ }\href
  {https://doi.org/10.1063/1.471928} {\bibfield  {journal} {\bibinfo  {journal}
  {The Journal of Chemical Physics}\ }\textbf {\bibinfo {volume} {105}},\
  \bibinfo {pages} {732} (\bibinfo {year} {1996})}\BibitemShut {NoStop}%
\bibitem [{\citenamefont {Mori}\ \emph {et~al.}(2020)\citenamefont {Mori},
  \citenamefont {Jiang}, \citenamefont {Fujii}, \citenamefont {Kitani},
  \citenamefont {Mizuno}, \citenamefont {Koreeda}, \citenamefont {Motoji},
  \citenamefont {Tokoro}, \citenamefont {Shiraki}, \citenamefont {Yamamoto},\
  and\ \citenamefont {Kojima}}]{Mori_2020}%
  \BibitemOpen
  \bibfield  {author} {\bibinfo {author} {\bibfnamefont {T.}~\bibnamefont
  {Mori}}, \bibinfo {author} {\bibfnamefont {Y.}~\bibnamefont {Jiang}},
  \bibinfo {author} {\bibfnamefont {Y.}~\bibnamefont {Fujii}}, \bibinfo
  {author} {\bibfnamefont {S.}~\bibnamefont {Kitani}}, \bibinfo {author}
  {\bibfnamefont {H.}~\bibnamefont {Mizuno}}, \bibinfo {author} {\bibfnamefont
  {A.}~\bibnamefont {Koreeda}}, \bibinfo {author} {\bibfnamefont
  {L.}~\bibnamefont {Motoji}}, \bibinfo {author} {\bibfnamefont
  {H.}~\bibnamefont {Tokoro}}, \bibinfo {author} {\bibfnamefont
  {K.}~\bibnamefont {Shiraki}}, \bibinfo {author} {\bibfnamefont
  {Y.}~\bibnamefont {Yamamoto}},\ and\ \bibinfo {author} {\bibfnamefont
  {S.}~\bibnamefont {Kojima}},\ }\href
  {https://doi.org/10.1103/PhysRevE.102.022502} {\bibfield  {journal} {\bibinfo
   {journal} {Physical Review E}\ }\textbf {\bibinfo {volume} {102}},\ \bibinfo
  {pages} {022502} (\bibinfo {year} {2020})}\BibitemShut {NoStop}%
\bibitem [{\citenamefont {Monnier}\ \emph {et~al.}(2021)\citenamefont
  {Monnier}, \citenamefont {Colmenero}, \citenamefont {Wolf},\ and\
  \citenamefont {Cangialosi}}]{Monnier_2021}%
  \BibitemOpen
  \bibfield  {author} {\bibinfo {author} {\bibfnamefont {X.}~\bibnamefont
  {Monnier}}, \bibinfo {author} {\bibfnamefont {J.}~\bibnamefont {Colmenero}},
  \bibinfo {author} {\bibfnamefont {M.}~\bibnamefont {Wolf}},\ and\ \bibinfo
  {author} {\bibfnamefont {D.}~\bibnamefont {Cangialosi}},\ }\href
  {https://doi.org/10.1103/physrevlett.126.118004} {\bibfield  {journal}
  {\bibinfo  {journal} {Physical Review Letters}\ }\textbf {\bibinfo {volume}
  {126}},\ \bibinfo {pages} {118004} (\bibinfo {year} {2021})}\BibitemShut
  {NoStop}%
\bibitem [{\citenamefont {Wyart}\ \emph {et~al.}(2005)\citenamefont {Wyart},
  \citenamefont {Silbert}, \citenamefont {Nagel},\ and\ \citenamefont
  {Witten}}]{Wyart_2006}%
  \BibitemOpen
  \bibfield  {author} {\bibinfo {author} {\bibfnamefont {M.}~\bibnamefont
  {Wyart}}, \bibinfo {author} {\bibfnamefont {L.~E.}\ \bibnamefont {Silbert}},
  \bibinfo {author} {\bibfnamefont {S.~R.}\ \bibnamefont {Nagel}},\ and\
  \bibinfo {author} {\bibfnamefont {T.~A.}\ \bibnamefont {Witten}},\ }\href
  {https://doi.org/10.1103/PhysRevE.72.051306} {\bibfield  {journal} {\bibinfo
  {journal} {Physical Review E}\ }\textbf {\bibinfo {volume} {72}},\ \bibinfo
  {pages} {051306} (\bibinfo {year} {2005})}\BibitemShut {NoStop}%
\bibitem [{\citenamefont {Xu}\ \emph {et~al.}(2007)\citenamefont {Xu},
  \citenamefont {Wyart}, \citenamefont {Liu},\ and\ \citenamefont
  {Nagel}}]{Xu_2007}%
  \BibitemOpen
  \bibfield  {author} {\bibinfo {author} {\bibfnamefont {N.}~\bibnamefont
  {Xu}}, \bibinfo {author} {\bibfnamefont {M.}~\bibnamefont {Wyart}}, \bibinfo
  {author} {\bibfnamefont {A.~J.}\ \bibnamefont {Liu}},\ and\ \bibinfo {author}
  {\bibfnamefont {S.~R.}\ \bibnamefont {Nagel}},\ }\href
  {https://doi.org/10.1103/physrevlett.98.175502} {\bibfield  {journal}
  {\bibinfo  {journal} {Physical Review Letters}\ }\textbf {\bibinfo {volume}
  {98}},\ \bibinfo {pages} {175502} (\bibinfo {year} {2007})}\BibitemShut
  {NoStop}%
\bibitem [{\citenamefont {Mizuno}\ \emph {et~al.}(2021)\citenamefont {Mizuno},
  \citenamefont {Hachiya},\ and\ \citenamefont {Ikeda}}]{Mizuno_2021}%
  \BibitemOpen
  \bibfield  {author} {\bibinfo {author} {\bibfnamefont {H.}~\bibnamefont
  {Mizuno}}, \bibinfo {author} {\bibfnamefont {M.}~\bibnamefont {Hachiya}},\
  and\ \bibinfo {author} {\bibfnamefont {A.}~\bibnamefont {Ikeda}},\ }\href
  {https://doi.org/10.1063/5.0072863} {\bibfield  {journal} {\bibinfo
  {journal} {The Journal of Chemical Physics}\ }\textbf {\bibinfo {volume}
  {155}},\ \bibinfo {pages} {234502} (\bibinfo {year} {2021})}\BibitemShut
  {NoStop}%
\bibitem [{\citenamefont {Lerner}\ and\ \citenamefont
  {Bouchbinder}(2022)}]{Lerner_2022}%
  \BibitemOpen
  \bibfield  {author} {\bibinfo {author} {\bibfnamefont {E.}~\bibnamefont
  {Lerner}}\ and\ \bibinfo {author} {\bibfnamefont {E.}~\bibnamefont
  {Bouchbinder}},\ }\href {https://arxiv.org/abs/2208.05725} {\bibfield
  {journal} {\bibinfo  {journal} {arXiv preprint arXiv:2208.05725}\ } (\bibinfo
  {year} {2022})}\BibitemShut {NoStop}%
\bibitem [{\citenamefont {Bhowmik}\ \emph {et~al.}(2019)\citenamefont
  {Bhowmik}, \citenamefont {Chaudhuri},\ and\ \citenamefont
  {Karmakar}}]{Bhowmik2019Effect}%
  \BibitemOpen
  \bibfield  {author} {\bibinfo {author} {\bibfnamefont {B.~P.}\ \bibnamefont
  {Bhowmik}}, \bibinfo {author} {\bibfnamefont {P.}~\bibnamefont {Chaudhuri}},\
  and\ \bibinfo {author} {\bibfnamefont {S.}~\bibnamefont {Karmakar}},\ }\href
  {https://doi.org/10.1103/PhysRevLett.123.185501} {\bibfield  {journal}
  {\bibinfo  {journal} {Physical Review Letters}\ }\textbf {\bibinfo {volume}
  {123}},\ \bibinfo {pages} {185501} (\bibinfo {year} {2019})}\BibitemShut
  {NoStop}%
\bibitem [{\citenamefont {Mizuno}\ and\ \citenamefont
  {Ikeda}(2018)}]{Mizuno_2018}%
  \BibitemOpen
  \bibfield  {author} {\bibinfo {author} {\bibfnamefont {H.}~\bibnamefont
  {Mizuno}}\ and\ \bibinfo {author} {\bibfnamefont {A.}~\bibnamefont {Ikeda}},\
  }\href {https://doi.org/10.1103/physreve.98.062612} {\bibfield  {journal}
  {\bibinfo  {journal} {Physical Review E}\ }\textbf {\bibinfo {volume} {98}},\
  \bibinfo {pages} {062612} (\bibinfo {year} {2018})}\BibitemShut {NoStop}%
\bibitem [{\citenamefont {Wijtmans}\ and\ \citenamefont
  {Manning}(2017)}]{Wijtmans_2017}%
  \BibitemOpen
  \bibfield  {author} {\bibinfo {author} {\bibfnamefont {S.}~\bibnamefont
  {Wijtmans}}\ and\ \bibinfo {author} {\bibfnamefont {M.~L.}\ \bibnamefont
  {Manning}},\ }\href {https://doi.org/10.1039/C7SM00792B} {\bibfield
  {journal} {\bibinfo  {journal} {Soft Matter}\ }\textbf {\bibinfo {volume}
  {13}},\ \bibinfo {pages} {5649} (\bibinfo {year} {2017})}\BibitemShut
  {NoStop}%
\bibitem [{\citenamefont {Richard}\ \emph {et~al.}(2020)\citenamefont
  {Richard}, \citenamefont {Gonz{\'a}lez-L{\'o}pez}, \citenamefont {Kapteijns},
  \citenamefont {Pater}, \citenamefont {Vaknin}, \citenamefont {Bouchbinder},\
  and\ \citenamefont {Lerner}}]{Richard_2020}%
  \BibitemOpen
  \bibfield  {author} {\bibinfo {author} {\bibfnamefont {D.}~\bibnamefont
  {Richard}}, \bibinfo {author} {\bibfnamefont {K.}~\bibnamefont
  {Gonz{\'a}lez-L{\'o}pez}}, \bibinfo {author} {\bibfnamefont {G.}~\bibnamefont
  {Kapteijns}}, \bibinfo {author} {\bibfnamefont {R.}~\bibnamefont {Pater}},
  \bibinfo {author} {\bibfnamefont {T.}~\bibnamefont {Vaknin}}, \bibinfo
  {author} {\bibfnamefont {E.}~\bibnamefont {Bouchbinder}},\ and\ \bibinfo
  {author} {\bibfnamefont {E.}~\bibnamefont {Lerner}},\ }\href
  {https://doi.org/10.1103/physrevlett.125.085502} {\bibfield  {journal}
  {\bibinfo  {journal} {Physical Review Letters}\ }\textbf {\bibinfo {volume}
  {125}},\ \bibinfo {pages} {085502} (\bibinfo {year} {2020})}\BibitemShut
  {NoStop}%
\bibitem [{\citenamefont {Bonfanti}\ \emph {et~al.}(2020)\citenamefont
  {Bonfanti}, \citenamefont {Guerra}, \citenamefont {Mondal}, \citenamefont
  {Procaccia},\ and\ \citenamefont {Zapperi}}]{Zapperi_2020}%
  \BibitemOpen
  \bibfield  {author} {\bibinfo {author} {\bibfnamefont {S.}~\bibnamefont
  {Bonfanti}}, \bibinfo {author} {\bibfnamefont {R.}~\bibnamefont {Guerra}},
  \bibinfo {author} {\bibfnamefont {C.}~\bibnamefont {Mondal}}, \bibinfo
  {author} {\bibfnamefont {I.}~\bibnamefont {Procaccia}},\ and\ \bibinfo
  {author} {\bibfnamefont {S.}~\bibnamefont {Zapperi}},\ }\href
  {https://doi.org/10.1103/physrevlett.125.085501} {\bibfield  {journal}
  {\bibinfo  {journal} {Physical Review Letters}\ }\textbf {\bibinfo {volume}
  {125}},\ \bibinfo {pages} {085501} (\bibinfo {year} {2020})}\BibitemShut
  {NoStop}%
\bibitem [{\citenamefont {Shiraishi}\ \emph {et~al.}(2020)\citenamefont
  {Shiraishi}, \citenamefont {Mizuno},\ and\ \citenamefont
  {Ikeda}}]{Shiraishi_2020}%
  \BibitemOpen
  \bibfield  {author} {\bibinfo {author} {\bibfnamefont {K.}~\bibnamefont
  {Shiraishi}}, \bibinfo {author} {\bibfnamefont {H.}~\bibnamefont {Mizuno}},\
  and\ \bibinfo {author} {\bibfnamefont {A.}~\bibnamefont {Ikeda}},\ }\href
  {https://doi.org/10.7566/JPSJ.89.074603} {\bibfield  {journal} {\bibinfo
  {journal} {Journal of the Physical Society of Japan}\ }\textbf {\bibinfo
  {volume} {89}},\ \bibinfo {pages} {074603} (\bibinfo {year}
  {2020})}\BibitemShut {NoStop}%
\bibitem [{\citenamefont {Gurarie}\ and\ \citenamefont
  {Chalker}(2003)}]{Gurarie_2003}%
  \BibitemOpen
  \bibfield  {author} {\bibinfo {author} {\bibfnamefont {V.}~\bibnamefont
  {Gurarie}}\ and\ \bibinfo {author} {\bibfnamefont {J.~T.}\ \bibnamefont
  {Chalker}},\ }\href {https://doi.org/10.1103/physrevb.68.134207} {\bibfield
  {journal} {\bibinfo  {journal} {Physical Review B}\ }\textbf {\bibinfo
  {volume} {68}},\ \bibinfo {pages} {134207} (\bibinfo {year}
  {2003})}\BibitemShut {NoStop}%
\bibitem [{\citenamefont {Gurevich}\ \emph {et~al.}(2003)\citenamefont
  {Gurevich}, \citenamefont {Parshin},\ and\ \citenamefont
  {Schober}}]{Gurevich_2003}%
  \BibitemOpen
  \bibfield  {author} {\bibinfo {author} {\bibfnamefont {V.~L.}\ \bibnamefont
  {Gurevich}}, \bibinfo {author} {\bibfnamefont {D.~A.}\ \bibnamefont
  {Parshin}},\ and\ \bibinfo {author} {\bibfnamefont {H.~R.}\ \bibnamefont
  {Schober}},\ }\href {https://doi.org/10.1103/physrevb.67.094203} {\bibfield
  {journal} {\bibinfo  {journal} {Physical Review B}\ }\textbf {\bibinfo
  {volume} {67}},\ \bibinfo {pages} {094203} (\bibinfo {year}
  {2003})}\BibitemShut {NoStop}%
\bibitem [{\citenamefont {Tanguy}\ \emph {et~al.}(2010)\citenamefont {Tanguy},
  \citenamefont {Mantisi},\ and\ \citenamefont {Tsamados}}]{Tanguy_2010}%
  \BibitemOpen
  \bibfield  {author} {\bibinfo {author} {\bibfnamefont {A.}~\bibnamefont
  {Tanguy}}, \bibinfo {author} {\bibfnamefont {B.}~\bibnamefont {Mantisi}},\
  and\ \bibinfo {author} {\bibfnamefont {M.}~\bibnamefont {Tsamados}},\ }\href
  {https://doi.org/10.1209/0295-5075/90/16004} {\bibfield  {journal} {\bibinfo
  {journal} {Europhysics Letters}\ }\textbf {\bibinfo {volume} {90}},\ \bibinfo
  {pages} {16004} (\bibinfo {year} {2010})}\BibitemShut {NoStop}%
\bibitem [{\citenamefont {Manning}\ and\ \citenamefont
  {Liu}(2011)}]{Manning_2011}%
  \BibitemOpen
  \bibfield  {author} {\bibinfo {author} {\bibfnamefont {M.~L.}\ \bibnamefont
  {Manning}}\ and\ \bibinfo {author} {\bibfnamefont {A.~J.}\ \bibnamefont
  {Liu}},\ }\href {https://doi.org/10.1103/PhysRevLett.107.108302} {\bibfield
  {journal} {\bibinfo  {journal} {Physical Review Letters}\ }\textbf {\bibinfo
  {volume} {107}},\ \bibinfo {pages} {108302} (\bibinfo {year}
  {2011})}\BibitemShut {NoStop}%
\bibitem [{\citenamefont {Oyama}\ \emph {et~al.}(2021)\citenamefont {Oyama},
  \citenamefont {Mizuno},\ and\ \citenamefont {Ikeda}}]{Oyama2021HB}%
  \BibitemOpen
  \bibfield  {author} {\bibinfo {author} {\bibfnamefont {N.}~\bibnamefont
  {Oyama}}, \bibinfo {author} {\bibfnamefont {H.}~\bibnamefont {Mizuno}},\ and\
  \bibinfo {author} {\bibfnamefont {A.}~\bibnamefont {Ikeda}},\ }\href
  {https://doi.org/10.1103/physrevlett.127.108003} {\bibfield  {journal}
  {\bibinfo  {journal} {Physical Review Letters}\ }\textbf {\bibinfo {volume}
  {127}},\ \bibinfo {pages} {108003} (\bibinfo {year} {2021})}\BibitemShut
  {NoStop}%
\bibitem [{\citenamefont {Widmer-Cooper}\ \emph {et~al.}(2008)\citenamefont
  {Widmer-Cooper}, \citenamefont {Perry}, \citenamefont {Harrowell},\ and\
  \citenamefont {Reichman}}]{Widmer_Cooper_2008}%
  \BibitemOpen
  \bibfield  {author} {\bibinfo {author} {\bibfnamefont {A.}~\bibnamefont
  {Widmer-Cooper}}, \bibinfo {author} {\bibfnamefont {H.}~\bibnamefont
  {Perry}}, \bibinfo {author} {\bibfnamefont {P.}~\bibnamefont {Harrowell}},\
  and\ \bibinfo {author} {\bibfnamefont {D.~R.}\ \bibnamefont {Reichman}},\
  }\href {https://doi.org/10.1038/nphys1025} {\bibfield  {journal} {\bibinfo
  {journal} {Nature Physics}\ }\textbf {\bibinfo {volume} {4}},\ \bibinfo
  {pages} {711} (\bibinfo {year} {2008})}\BibitemShut {NoStop}%
\bibitem [{\citenamefont {Lema{\^\i}tre}\ and\ \citenamefont
  {Maloney}(2006)}]{Lemaitre_2006}%
  \BibitemOpen
  \bibfield  {author} {\bibinfo {author} {\bibfnamefont {A.}~\bibnamefont
  {Lema{\^\i}tre}}\ and\ \bibinfo {author} {\bibfnamefont {C.}~\bibnamefont
  {Maloney}},\ }\href {https://doi.org/10.1007/s10955-005-9015-5} {\bibfield
  {journal} {\bibinfo  {journal} {Journal of Statistical Physics}\ }\textbf
  {\bibinfo {volume} {123}},\ \bibinfo {pages} {415} (\bibinfo {year}
  {2006})}\BibitemShut {NoStop}%
\bibitem [{\citenamefont {Mizuno}\ \emph {et~al.}(2016)\citenamefont {Mizuno},
  \citenamefont {Saitoh},\ and\ \citenamefont
  {Silbert}}]{Mizuno_Saitoh_Silbert_2016}%
  \BibitemOpen
  \bibfield  {author} {\bibinfo {author} {\bibfnamefont {H.}~\bibnamefont
  {Mizuno}}, \bibinfo {author} {\bibfnamefont {K.}~\bibnamefont {Saitoh}},\
  and\ \bibinfo {author} {\bibfnamefont {L.~E.}\ \bibnamefont {Silbert}},\
  }\href {https://doi.org/10.1103/physreve.93.062905} {\bibfield  {journal}
  {\bibinfo  {journal} {Physical Review E}\ }\textbf {\bibinfo {volume} {93}},\
  \bibinfo {pages} {062905} (\bibinfo {year} {2016})}\BibitemShut {NoStop}%
\bibitem [{\citenamefont {Anderson}\ \emph {et~al.}(1972)\citenamefont
  {Anderson}, \citenamefont {Halperin},\ and\ \citenamefont
  {Varma}}]{Anderson_1972}%
  \BibitemOpen
  \bibfield  {author} {\bibinfo {author} {\bibfnamefont {P.~W.}\ \bibnamefont
  {Anderson}}, \bibinfo {author} {\bibfnamefont {B.~I.}\ \bibnamefont
  {Halperin}},\ and\ \bibinfo {author} {\bibfnamefont {C.~M.}\ \bibnamefont
  {Varma}},\ }\href {https://doi.org/10.1080/14786437208229210} {\bibfield
  {journal} {\bibinfo  {journal} {Philosophical Magazine}\ }\textbf {\bibinfo
  {volume} {25}},\ \bibinfo {pages} {1} (\bibinfo {year} {1972})}\BibitemShut
  {NoStop}%
\bibitem [{\citenamefont {Phillips}(1987)}]{Phillips_1987}%
  \BibitemOpen
  \bibfield  {author} {\bibinfo {author} {\bibfnamefont {W.~A.}\ \bibnamefont
  {Phillips}},\ }\href {https://doi.org/10.1088/0034-4885/50/12/003} {\bibfield
   {journal} {\bibinfo  {journal} {Reports on Progress in Physics}\ }\textbf
  {\bibinfo {volume} {50}},\ \bibinfo {pages} {1657} (\bibinfo {year}
  {1987})}\BibitemShut {NoStop}%
\bibitem [{\citenamefont {Khomenko}\ \emph {et~al.}(2020)\citenamefont
  {Khomenko}, \citenamefont {Scalliet}, \citenamefont {Berthier}, \citenamefont
  {Reichman},\ and\ \citenamefont {Zamponi}}]{Khomenko_2020}%
  \BibitemOpen
  \bibfield  {author} {\bibinfo {author} {\bibfnamefont {D.}~\bibnamefont
  {Khomenko}}, \bibinfo {author} {\bibfnamefont {C.}~\bibnamefont {Scalliet}},
  \bibinfo {author} {\bibfnamefont {L.}~\bibnamefont {Berthier}}, \bibinfo
  {author} {\bibfnamefont {D.~R.}\ \bibnamefont {Reichman}},\ and\ \bibinfo
  {author} {\bibfnamefont {F.}~\bibnamefont {Zamponi}},\ }\href
  {https://doi.org/10.1103/PhysRevLett.124.225901} {\bibfield  {journal}
  {\bibinfo  {journal} {Physical Review Letters}\ }\textbf {\bibinfo {volume}
  {124}},\ \bibinfo {pages} {225901} (\bibinfo {year} {2020})}\BibitemShut
  {NoStop}%
\end{thebibliography}%


\begin{thebibliography}{10}%
\makeatletter
\providecommand \@ifxundefined [1]{%
 \@ifx{#1\undefined}
}%
\providecommand \@ifnum [1]{%
 \ifnum #1\expandafter \@firstoftwo
 \else \expandafter \@secondoftwo
 \fi
}%
\providecommand \@ifx [1]{%
 \ifx #1\expandafter \@firstoftwo
 \else \expandafter \@secondoftwo
 \fi
}%
\providecommand \natexlab [1]{#1}%
\providecommand \enquote  [1]{``#1''}%
\providecommand \bibnamefont  [1]{#1}%
\providecommand \bibfnamefont [1]{#1}%
\providecommand \citenamefont [1]{#1}%
\providecommand \href@noop [0]{\@secondoftwo}%
\providecommand \href [0]{\begingroup \@sanitize@url \@href}%
\providecommand \@href[1]{\@@startlink{#1}\@@href}%
\providecommand \@@href[1]{\endgroup#1\@@endlink}%
\providecommand \@sanitize@url [0]{\catcode `\\12\catcode `\$12\catcode
  `\&12\catcode `\#12\catcode `\^12\catcode `\_12\catcode `\%12\relax}%
\providecommand \@@startlink[1]{}%
\providecommand \@@endlink[0]{}%
\providecommand \url  [0]{\begingroup\@sanitize@url \@url }%
\providecommand \@url [1]{\endgroup\@href {#1}{\urlprefix }}%
\providecommand \urlprefix  [0]{URL }%
\providecommand \Eprint [0]{\href }%
\providecommand \doibase [0]{https://doi.org/}%
\providecommand \selectlanguage [0]{\@gobble}%
\providecommand \bibinfo  [0]{\@secondoftwo}%
\providecommand \bibfield  [0]{\@secondoftwo}%
\providecommand \translation [1]{[#1]}%
\providecommand \BibitemOpen [0]{}%
\providecommand \bibitemStop [0]{}%
\providecommand \bibitemNoStop [0]{.\EOS\space}%
\providecommand \EOS [0]{\spacefactor3000\relax}%
\providecommand \BibitemShut  [1]{\csname bibitem#1\endcsname}%
\let\auto@bib@innerbib\@empty
\bibitem [{\citenamefont {Kob}\ and\ \citenamefont
  {Andersen}(1994)}]{Kob_Andersen_PRL_1994}%
  \BibitemOpen
  \bibfield  {author} {\bibinfo {author} {\bibfnamefont {W.}~\bibnamefont
  {Kob}}\ and\ \bibinfo {author} {\bibfnamefont {H.~C.}\ \bibnamefont
  {Andersen}},\ }\href {https://doi.org/10.1103/PhysRevLett.73.1376} {\bibfield
   {journal} {\bibinfo  {journal} {Physical Review Letters}\ }\textbf {\bibinfo
  {volume} {73}},\ \bibinfo {pages} {1376} (\bibinfo {year}
  {1994})}\BibitemShut {NoStop}%
\bibitem [{\citenamefont {Kob}\ and\ \citenamefont
  {Andersen}(1995)}]{Kob_Andersen_I_1995}%
  \BibitemOpen
  \bibfield  {author} {\bibinfo {author} {\bibfnamefont {W.}~\bibnamefont
  {Kob}}\ and\ \bibinfo {author} {\bibfnamefont {H.~C.}\ \bibnamefont
  {Andersen}},\ }\href {https://doi.org/10.1103/PhysRevE.51.4626} {\bibfield
  {journal} {\bibinfo  {journal} {Physical Review E}\ }\textbf {\bibinfo
  {volume} {51}},\ \bibinfo {pages} {4626} (\bibinfo {year}
  {1995})}\BibitemShut {NoStop}%
\bibitem [{\citenamefont {Shimada}\ \emph {et~al.}(2018)\citenamefont
  {Shimada}, \citenamefont {Mizuno},\ and\ \citenamefont
  {Ikeda}}]{Shimada_LJ_2018}%
  \BibitemOpen
  \bibfield  {author} {\bibinfo {author} {\bibfnamefont {M.}~\bibnamefont
  {Shimada}}, \bibinfo {author} {\bibfnamefont {H.}~\bibnamefont {Mizuno}},\
  and\ \bibinfo {author} {\bibfnamefont {A.}~\bibnamefont {Ikeda}},\ }\href
  {https://doi.org/10.1103/physreve.97.022609} {\bibfield  {journal} {\bibinfo
  {journal} {Physical Review E}\ }\textbf {\bibinfo {volume} {97}},\ \bibinfo
  {pages} {022609} (\bibinfo {year} {2018})}\BibitemShut {NoStop}%
\bibitem [{\citenamefont {Allen}\ and\ \citenamefont
  {Tildesley}(2017)}]{Allen_Tildesley}%
  \BibitemOpen
  \bibfield  {author} {\bibinfo {author} {\bibfnamefont {M.~P.}\ \bibnamefont
  {Allen}}\ and\ \bibinfo {author} {\bibfnamefont {D.~J.}\ \bibnamefont
  {Tildesley}},\ }\href {https://doi.org/10.1093/oso/9780198803195.001.0001}
  {\emph {\bibinfo {title} {{Computer Simulation of Liquids}}}}\ (\bibinfo
  {publisher} {Oxford University Press},\ \bibinfo {year} {2017})\BibitemShut
  {NoStop}%
\bibitem [{\citenamefont {Frenkel}\ and\ \citenamefont
  {Smit}(2002)}]{Frenkel_Smit}%
  \BibitemOpen
  \bibfield  {author} {\bibinfo {author} {\bibfnamefont {D.}~\bibnamefont
  {Frenkel}}\ and\ \bibinfo {author} {\bibfnamefont {B.}~\bibnamefont {Smit}},\
  }\href {https://doi.org/10.1016/B978-0-12-267351-1.X5000-7} {\emph {\bibinfo
  {title} {{Understanding Molecular Simulation}}}}\ (\bibinfo  {publisher}
  {Academic Press},\ \bibinfo {year} {2002})\BibitemShut {NoStop}%
\bibitem [{\citenamefont {Kob}\ and\ \citenamefont
  {Berthier}(2013)}]{Kob_2013}%
  \BibitemOpen
  \bibfield  {author} {\bibinfo {author} {\bibfnamefont {W.}~\bibnamefont
  {Kob}}\ and\ \bibinfo {author} {\bibfnamefont {L.}~\bibnamefont {Berthier}},\
  }\href {https://doi.org/10.1103/physrevlett.110.245702} {\bibfield  {journal}
  {\bibinfo  {journal} {Physical Review Letters}\ }\textbf {\bibinfo {volume}
  {110}},\ \bibinfo {pages} {245702} (\bibinfo {year} {2013})}\BibitemShut
  {NoStop}%
\bibitem [{\citenamefont {Ozawa}\ \emph {et~al.}(2015)\citenamefont {Ozawa},
  \citenamefont {Kob}, \citenamefont {Ikeda},\ and\ \citenamefont
  {Miyazaki}}]{Ozawa_2015}%
  \BibitemOpen
  \bibfield  {author} {\bibinfo {author} {\bibfnamefont {M.}~\bibnamefont
  {Ozawa}}, \bibinfo {author} {\bibfnamefont {W.}~\bibnamefont {Kob}}, \bibinfo
  {author} {\bibfnamefont {A.}~\bibnamefont {Ikeda}},\ and\ \bibinfo {author}
  {\bibfnamefont {K.}~\bibnamefont {Miyazaki}},\ }\href
  {https://doi.org/10.1073/pnas.1500730112} {\bibfield  {journal} {\bibinfo
  {journal} {Proceedings of the National Academy of Sciences}\ }\textbf
  {\bibinfo {volume} {112}},\ \bibinfo {pages} {6914} (\bibinfo {year}
  {2015})}\BibitemShut {NoStop}%
\bibitem [{\citenamefont {Gu{\'e}nol{\'e}}\ \emph {et~al.}(2020)\citenamefont
  {Gu{\'e}nol{\'e}}, \citenamefont {N{\"o}hring}, \citenamefont {Vaid},
  \citenamefont {Houll{\'e}}, \citenamefont {Xie}, \citenamefont {Prakash},\
  and\ \citenamefont {Bitzek}}]{Guenole_2020}%
  \BibitemOpen
  \bibfield  {author} {\bibinfo {author} {\bibfnamefont {J.}~\bibnamefont
  {Gu{\'e}nol{\'e}}}, \bibinfo {author} {\bibfnamefont {W.~G.}\ \bibnamefont
  {N{\"o}hring}}, \bibinfo {author} {\bibfnamefont {A.}~\bibnamefont {Vaid}},
  \bibinfo {author} {\bibfnamefont {F.}~\bibnamefont {Houll{\'e}}}, \bibinfo
  {author} {\bibfnamefont {Z.}~\bibnamefont {Xie}}, \bibinfo {author}
  {\bibfnamefont {A.}~\bibnamefont {Prakash}},\ and\ \bibinfo {author}
  {\bibfnamefont {E.}~\bibnamefont {Bitzek}},\ }\href
  {https://doi.org/10.1016/j.commatsci.2020.109584} {\bibfield  {journal}
  {\bibinfo  {journal} {Computational Materials Science}\ }\textbf {\bibinfo
  {volume} {175}},\ \bibinfo {pages} {109584} (\bibinfo {year}
  {2020})}\BibitemShut {NoStop}%
\bibitem [{\citenamefont {Guennebaud}\ \emph {et~al.}(2010)\citenamefont
  {Guennebaud}, \citenamefont {Jacob} \emph {et~al.}}]{eigenweb}%
  \BibitemOpen
  \bibfield  {author} {\bibinfo {author} {\bibfnamefont {G.}~\bibnamefont
  {Guennebaud}}, \bibinfo {author} {\bibfnamefont {B.}~\bibnamefont {Jacob}},
  \emph {et~al.},\ }\href@noop {} {\bibinfo {title} {Eigen v3}},\ \bibinfo
  {howpublished} {http://eigen.tuxfamily.org} (\bibinfo {year}
  {2010})\BibitemShut {NoStop}%
\bibitem [{\citenamefont {Virtanen}\ \emph {et~al.}(2020)\citenamefont
  {Virtanen}, \citenamefont {Gommers}, \citenamefont {Oliphant}, \citenamefont
  {Haberland}, \citenamefont {Reddy}, \citenamefont {Cournapeau}, \citenamefont
  {Burovski}, \citenamefont {Peterson}, \citenamefont {Weckesser},
  \citenamefont {Bright}, \citenamefont {van~der Walt}, \citenamefont {Brett},
  \citenamefont {Wilson}, \citenamefont {Millman}, \citenamefont {Mayorov},
  \citenamefont {Nelson}, \citenamefont {Jones}, \citenamefont {Kern},
  \citenamefont {Larson}, \citenamefont {Carey}, \citenamefont {Polat},
  \citenamefont {Feng}, \citenamefont {Moore}, \citenamefont {VanderPlas},
  \citenamefont {Laxalde}, \citenamefont {Perktold}, \citenamefont {Cimrman},
  \citenamefont {Henriksen}, \citenamefont {Quintero}, \citenamefont {Harris},
  \citenamefont {Archibald}, \citenamefont {Ribeiro}, \citenamefont
  {Pedregosa}, \citenamefont {van Mulbregt}, \citenamefont {Vijaykumar},
  \citenamefont {Bardelli}, \citenamefont {Rothberg}, \citenamefont {Hilboll},
  \citenamefont {Kloeckner}, \citenamefont {Scopatz}, \citenamefont {Lee},
  \citenamefont {Rokem}, \citenamefont {Woods}, \citenamefont {Fulton},
  \citenamefont {Masson}, \citenamefont {H{\"a}ggstr{\"o}m}, \citenamefont
  {Fitzgerald}, \citenamefont {Nicholson}, \citenamefont {Hagen}, \citenamefont
  {Pasechnik}, \citenamefont {Olivetti}, \citenamefont {Martin}, \citenamefont
  {Wieser}, \citenamefont {Silva}, \citenamefont {Lenders}, \citenamefont
  {Wilhelm}, \citenamefont {Young}, \citenamefont {Price}, \citenamefont
  {Ingold}, \citenamefont {Allen}, \citenamefont {Lee}, \citenamefont {Audren},
  \citenamefont {Probst}, \citenamefont {Dietrich}, \citenamefont {Silterra},
  \citenamefont {Webber}, \citenamefont {Slavi{\v{c}}}, \citenamefont
  {Nothman}, \citenamefont {Buchner}, \citenamefont {Kulick}, \citenamefont
  {Sch{\"o}nberger}, \citenamefont {de~Miranda~Cardoso}, \citenamefont
  {Reimer}, \citenamefont {Harrington}, \citenamefont {Rodr{\'i}guez},
  \citenamefont {Nunez-Iglesias}, \citenamefont {Kuczynski}, \citenamefont
  {Tritz}, \citenamefont {Thoma}, \citenamefont {Newville}, \citenamefont
  {K{\"u}mmerer}, \citenamefont {Bolingbroke}, \citenamefont {Tartre},
  \citenamefont {Pak}, \citenamefont {Smith}, \citenamefont {Nowaczyk},
  \citenamefont {Shebanov}, \citenamefont {Pavlyk}, \citenamefont {Brodtkorb},
  \citenamefont {Lee}, \citenamefont {McGibbon}, \citenamefont {Feldbauer},
  \citenamefont {Lewis}, \citenamefont {Tygier}, \citenamefont {Sievert},
  \citenamefont {Vigna}, \citenamefont {Peterson}, \citenamefont {More},
  \citenamefont {Pudlik}, \citenamefont {Oshima}, \citenamefont {Pingel},
  \citenamefont {Robitaille}, \citenamefont {Spura}, \citenamefont {Jones},
  \citenamefont {Cera}, \citenamefont {Leslie}, \citenamefont {Zito},
  \citenamefont {Krauss}, \citenamefont {Upadhyay}, \citenamefont {Halchenko},
  \citenamefont {V{\'a}zquez-Baeza},\ and\ \citenamefont {{SciPy 1.0
  Contributors}}}]{Virtanen2020}%
  \BibitemOpen
  \bibfield  {author} {\bibinfo {author} {\bibfnamefont {P.}~\bibnamefont
  {Virtanen}}, \bibinfo {author} {\bibfnamefont {R.}~\bibnamefont {Gommers}},
  \bibinfo {author} {\bibfnamefont {T.~E.}\ \bibnamefont {Oliphant}}, \bibinfo
  {author} {\bibfnamefont {M.}~\bibnamefont {Haberland}}, \bibinfo {author}
  {\bibfnamefont {T.}~\bibnamefont {Reddy}}, \bibinfo {author} {\bibfnamefont
  {D.}~\bibnamefont {Cournapeau}}, \bibinfo {author} {\bibfnamefont
  {E.}~\bibnamefont {Burovski}}, \bibinfo {author} {\bibfnamefont
  {P.}~\bibnamefont {Peterson}}, \bibinfo {author} {\bibfnamefont
  {W.}~\bibnamefont {Weckesser}}, \bibinfo {author} {\bibfnamefont
  {J.}~\bibnamefont {Bright}}, \bibinfo {author} {\bibfnamefont {S.~J.}\
  \bibnamefont {van~der Walt}}, \bibinfo {author} {\bibfnamefont
  {M.}~\bibnamefont {Brett}}, \bibinfo {author} {\bibfnamefont
  {J.}~\bibnamefont {Wilson}}, \bibinfo {author} {\bibfnamefont {K.~J.}\
  \bibnamefont {Millman}}, \bibinfo {author} {\bibfnamefont {N.}~\bibnamefont
  {Mayorov}}, \bibinfo {author} {\bibfnamefont {A.~R.~J.}\ \bibnamefont
  {Nelson}}, \bibinfo {author} {\bibfnamefont {E.}~\bibnamefont {Jones}},
  \bibinfo {author} {\bibfnamefont {R.}~\bibnamefont {Kern}}, \bibinfo {author}
  {\bibfnamefont {E.}~\bibnamefont {Larson}}, \bibinfo {author} {\bibfnamefont
  {C.~J.}\ \bibnamefont {Carey}}, \bibinfo {author} {\bibfnamefont
  {{\.{I}}.}~\bibnamefont {Polat}}, \bibinfo {author} {\bibfnamefont
  {Y.}~\bibnamefont {Feng}}, \bibinfo {author} {\bibfnamefont {E.~W.}\
  \bibnamefont {Moore}}, \bibinfo {author} {\bibfnamefont {J.}~\bibnamefont
  {VanderPlas}}, \bibinfo {author} {\bibfnamefont {D.}~\bibnamefont {Laxalde}},
  \bibinfo {author} {\bibfnamefont {J.}~\bibnamefont {Perktold}}, \bibinfo
  {author} {\bibfnamefont {R.}~\bibnamefont {Cimrman}}, \bibinfo {author}
  {\bibfnamefont {I.}~\bibnamefont {Henriksen}}, \bibinfo {author}
  {\bibfnamefont {E.~A.}\ \bibnamefont {Quintero}}, \bibinfo {author}
  {\bibfnamefont {C.~R.}\ \bibnamefont {Harris}}, \bibinfo {author}
  {\bibfnamefont {A.~M.}\ \bibnamefont {Archibald}}, \bibinfo {author}
  {\bibfnamefont {A.~H.}\ \bibnamefont {Ribeiro}}, \bibinfo {author}
  {\bibfnamefont {F.}~\bibnamefont {Pedregosa}}, \bibinfo {author}
  {\bibfnamefont {P.}~\bibnamefont {van Mulbregt}}, \bibinfo {author}
  {\bibfnamefont {A.}~\bibnamefont {Vijaykumar}}, \bibinfo {author}
  {\bibfnamefont {A.~P.}\ \bibnamefont {Bardelli}}, \bibinfo {author}
  {\bibfnamefont {A.}~\bibnamefont {Rothberg}}, \bibinfo {author}
  {\bibfnamefont {A.}~\bibnamefont {Hilboll}}, \bibinfo {author} {\bibfnamefont
  {A.}~\bibnamefont {Kloeckner}}, \bibinfo {author} {\bibfnamefont
  {A.}~\bibnamefont {Scopatz}}, \bibinfo {author} {\bibfnamefont
  {A.}~\bibnamefont {Lee}}, \bibinfo {author} {\bibfnamefont {A.}~\bibnamefont
  {Rokem}}, \bibinfo {author} {\bibfnamefont {C.~N.}\ \bibnamefont {Woods}},
  \bibinfo {author} {\bibfnamefont {C.}~\bibnamefont {Fulton}}, \bibinfo
  {author} {\bibfnamefont {C.}~\bibnamefont {Masson}}, \bibinfo {author}
  {\bibfnamefont {C.}~\bibnamefont {H{\"a}ggstr{\"o}m}}, \bibinfo {author}
  {\bibfnamefont {C.}~\bibnamefont {Fitzgerald}}, \bibinfo {author}
  {\bibfnamefont {D.~A.}\ \bibnamefont {Nicholson}}, \bibinfo {author}
  {\bibfnamefont {D.~R.}\ \bibnamefont {Hagen}}, \bibinfo {author}
  {\bibfnamefont {D.~V.}\ \bibnamefont {Pasechnik}}, \bibinfo {author}
  {\bibfnamefont {E.}~\bibnamefont {Olivetti}}, \bibinfo {author}
  {\bibfnamefont {E.}~\bibnamefont {Martin}}, \bibinfo {author} {\bibfnamefont
  {E.}~\bibnamefont {Wieser}}, \bibinfo {author} {\bibfnamefont
  {F.}~\bibnamefont {Silva}}, \bibinfo {author} {\bibfnamefont
  {F.}~\bibnamefont {Lenders}}, \bibinfo {author} {\bibfnamefont
  {F.}~\bibnamefont {Wilhelm}}, \bibinfo {author} {\bibfnamefont
  {G.}~\bibnamefont {Young}}, \bibinfo {author} {\bibfnamefont {G.~A.}\
  \bibnamefont {Price}}, \bibinfo {author} {\bibfnamefont {G.-L.}\ \bibnamefont
  {Ingold}}, \bibinfo {author} {\bibfnamefont {G.~E.}\ \bibnamefont {Allen}},
  \bibinfo {author} {\bibfnamefont {G.~R.}\ \bibnamefont {Lee}}, \bibinfo
  {author} {\bibfnamefont {H.}~\bibnamefont {Audren}}, \bibinfo {author}
  {\bibfnamefont {I.}~\bibnamefont {Probst}}, \bibinfo {author} {\bibfnamefont
  {J.~P.}\ \bibnamefont {Dietrich}}, \bibinfo {author} {\bibfnamefont
  {J.}~\bibnamefont {Silterra}}, \bibinfo {author} {\bibfnamefont {J.~T.}\
  \bibnamefont {Webber}}, \bibinfo {author} {\bibfnamefont {J.}~\bibnamefont
  {Slavi{\v{c}}}}, \bibinfo {author} {\bibfnamefont {J.}~\bibnamefont
  {Nothman}}, \bibinfo {author} {\bibfnamefont {J.}~\bibnamefont {Buchner}},
  \bibinfo {author} {\bibfnamefont {J.}~\bibnamefont {Kulick}}, \bibinfo
  {author} {\bibfnamefont {J.~L.}\ \bibnamefont {Sch{\"o}nberger}}, \bibinfo
  {author} {\bibfnamefont {J.~V.}\ \bibnamefont {de~Miranda~Cardoso}}, \bibinfo
  {author} {\bibfnamefont {J.}~\bibnamefont {Reimer}}, \bibinfo {author}
  {\bibfnamefont {J.}~\bibnamefont {Harrington}}, \bibinfo {author}
  {\bibfnamefont {J.~L.~C.}\ \bibnamefont {Rodr{\'i}guez}}, \bibinfo {author}
  {\bibfnamefont {J.}~\bibnamefont {Nunez-Iglesias}}, \bibinfo {author}
  {\bibfnamefont {J.}~\bibnamefont {Kuczynski}}, \bibinfo {author}
  {\bibfnamefont {K.}~\bibnamefont {Tritz}}, \bibinfo {author} {\bibfnamefont
  {M.}~\bibnamefont {Thoma}}, \bibinfo {author} {\bibfnamefont
  {M.}~\bibnamefont {Newville}}, \bibinfo {author} {\bibfnamefont
  {M.}~\bibnamefont {K{\"u}mmerer}}, \bibinfo {author} {\bibfnamefont
  {M.}~\bibnamefont {Bolingbroke}}, \bibinfo {author} {\bibfnamefont
  {M.}~\bibnamefont {Tartre}}, \bibinfo {author} {\bibfnamefont
  {M.}~\bibnamefont {Pak}}, \bibinfo {author} {\bibfnamefont {N.~J.}\
  \bibnamefont {Smith}}, \bibinfo {author} {\bibfnamefont {N.}~\bibnamefont
  {Nowaczyk}}, \bibinfo {author} {\bibfnamefont {N.}~\bibnamefont {Shebanov}},
  \bibinfo {author} {\bibfnamefont {O.}~\bibnamefont {Pavlyk}}, \bibinfo
  {author} {\bibfnamefont {P.~A.}\ \bibnamefont {Brodtkorb}}, \bibinfo {author}
  {\bibfnamefont {P.}~\bibnamefont {Lee}}, \bibinfo {author} {\bibfnamefont
  {R.~T.}\ \bibnamefont {McGibbon}}, \bibinfo {author} {\bibfnamefont
  {R.}~\bibnamefont {Feldbauer}}, \bibinfo {author} {\bibfnamefont
  {S.}~\bibnamefont {Lewis}}, \bibinfo {author} {\bibfnamefont
  {S.}~\bibnamefont {Tygier}}, \bibinfo {author} {\bibfnamefont
  {S.}~\bibnamefont {Sievert}}, \bibinfo {author} {\bibfnamefont
  {S.}~\bibnamefont {Vigna}}, \bibinfo {author} {\bibfnamefont
  {S.}~\bibnamefont {Peterson}}, \bibinfo {author} {\bibfnamefont
  {S.}~\bibnamefont {More}}, \bibinfo {author} {\bibfnamefont {T.}~\bibnamefont
  {Pudlik}}, \bibinfo {author} {\bibfnamefont {T.}~\bibnamefont {Oshima}},
  \bibinfo {author} {\bibfnamefont {T.~J.}\ \bibnamefont {Pingel}}, \bibinfo
  {author} {\bibfnamefont {T.~P.}\ \bibnamefont {Robitaille}}, \bibinfo
  {author} {\bibfnamefont {T.}~\bibnamefont {Spura}}, \bibinfo {author}
  {\bibfnamefont {T.~R.}\ \bibnamefont {Jones}}, \bibinfo {author}
  {\bibfnamefont {T.}~\bibnamefont {Cera}}, \bibinfo {author} {\bibfnamefont
  {T.}~\bibnamefont {Leslie}}, \bibinfo {author} {\bibfnamefont
  {T.}~\bibnamefont {Zito}}, \bibinfo {author} {\bibfnamefont {T.}~\bibnamefont
  {Krauss}}, \bibinfo {author} {\bibfnamefont {U.}~\bibnamefont {Upadhyay}},
  \bibinfo {author} {\bibfnamefont {Y.~O.}\ \bibnamefont {Halchenko}}, \bibinfo
  {author} {\bibfnamefont {Y.}~\bibnamefont {V{\'a}zquez-Baeza}},\ and\
  \bibinfo {author} {\bibnamefont {{SciPy 1.0 Contributors}}},\ }\href
  {https://doi.org/10.1038/s41592-019-0686-2} {\bibfield  {journal} {\bibinfo
  {journal} {Nature Methods}\ }\textbf {\bibinfo {volume} {17}},\ \bibinfo
  {pages} {261} (\bibinfo {year} {2020})}\BibitemShut {NoStop}%
\end{thebibliography}%
\end{document}


\title{Supplemental Material for Low-Frequency Vibrational States in Ideal Glasses with Random Pinning}
\author{Kumpei Shiraishi}
\email{kumpeishiraishi@g.ecc.u-tokyo.ac.jp}
\author{Yusuke Hara}
\author{Hideyuki Mizuno}
\affiliation{Graduate School of Arts and Sciences, University of Tokyo, Komaba, Tokyo 153-8902, Japan}
\date{\today}
\maketitle

\section{Model}
We consider the Kob-Andersen (KA) model~\cite{Kob_Andersen_PRL_1994,Kob_Andersen_I_1995} in three-dimensional space ($d=3$).
Each particle interacts with the Lennard-Jones potential
\begin{align}
 \phi(r_{ij}) = 4\epsilon_{ij}\bqty{\pqty{\frac{\sigma_{ij}}{r_{ij}}}^{12} - \pqty{\frac{\sigma_{ij}}{r_{ij}}}^6}.
\end{align}
Since the continuity of the pair force strongly affects the properties of modes at low frequencies~\cite{Shimada_LJ_2018}, we employ the force-shifted potential
\begin{align}
 V(r_{ij}) = \phi(r_{ij}) - \phi(r^\text{cut}_{ij}) - \phi^\prime(r^\text{cut}_{ij}) (r - r^\text{cut}_{ij}),
\end{align}
where $r^\text{cut}_{ij} = 2.5\sigma_{ij}$.
Both types of particles (A and B) have the same mass $m$, which we set to unity. The interaction parameters are chosen as follows:
$\sigma_\text{AA} = 1.0, \sigma_\text{AB} = 0.8, \sigma_\text{BB} = 0.88, \epsilon_\text{AA} = 1.0, \epsilon_\text{AB} = 1.5, \epsilon_\text{BB} = 0.5$.
Particles A and B are mixed in a ratio of 80:20.
Particles are enclosed in a square box with periodic boundary conditions. The linear size $L$ of the box is determined by the number density $\rho = 1.204$.
Lengths, energies, and time are measured in units of $\sigma_\text{AA}$, $\epsilon_\text{AA}$, and $\pqty{m\sigma_\text{AA}/\epsilon_\text{AA}}^{1/2}$, respectively.
The Boltzmann constant $k_\text{B}$ is set to unity.

\section{Preparation of equilibrium configurations}
We generate equilibrium configurations at $T_p = 0.45$ by molecular dynamics (MD) simulations using in-house code.
Starting from the equilibrium configurations at the onset temperature $T_o = 1.0$, we run MD simulations in the \textit{NVT} ensemble using the Nos\'{e}-Hoover thermostat~\cite{Allen_Tildesley,Frenkel_Smit} for 50 times the relaxation time $\tau_\alpha$ at $T_p$.
The time step of MD is $\Delta t = 0.005$.
For the case of $N = 1000$, we perform this procedure independently to sample the configurations at $T_p$.
For the case of $N = 4000$, we continue the MD simulation and sample the configurations after each $2\tau_\alpha$ elapses, reaching more than $1000\tau_\alpha$ as the total simulation length.
We perform 48 independent MD runs for the case of $N = 4000$.
$\tau_\alpha$ is defined by the self-intermediate scattering function as $F_s(k, \tau_\alpha) = e^{-1}$ with $k = 7.25$, corresponding to the peak of the structure factor~\cite{Kob_Andersen_PRL_1994}.
We call the equilibrium (unpinned) configurations parent configurations.

For configurations with lower numbers of particles that are used as ``templates,'' the same MD simulations of the KA system are performed at the onset temperature $T_o = 1.0$ for the same number of configurations as above~\cite{Kob_2013,Ozawa_2015}.
To select pinned particles in the parent configurations, we first rescale the positions of particles in ``template'' to set the linear size of the simulation box.
We sweep all particles in the parent configuration to find the closest one for each particle in the ``template.''
The $cN$ particles selected by this procedure are the pinned particles of the parent configuration.

\section{Energy minimization and vibrational analysis with pinned particles}
To obtain inherent structures, we use the FIRE algorithm~\cite{Guenole_2020}.
Since the positions of the particles do not change if the forces acting on them are zero, we fill the forces of the pinned particles with zero after the standard calculation of pair forces~\cite{Allen_Tildesley}.
The convergence of the algorithm is judged by whether the maximum value of the norms of the forces acting on each (unpinned) particle is less than $1.5 \times 10^{-12}$.

After energy minimization, we calculate the dynamical matrix $\mathcal{M}$, a real symmetric matrix whose size is $d(1-c)N \times d(1-c)N$.
Let $P$ be the set of pinned particle indices.
Suppose $i \notin P$; then, the diagonal part of $\mathcal{M}$ is
\begin{align}
 \mathcal{M}_{ii} = \sum_{j \in P}\pdv{V}{\vb*{r}_i}{\vb*{r}_j} + \sum^N_{\substack{j=1 \\ j \neq i \\ j \notin P}}\pdv{V}{\vb*{r}_i}{\vb*{r}_j},
\end{align}
where $V = \sum_{i,j} V(r_{ij})$ is the potential of the system.
The off-diagonal part of $\mathcal{M}$ is
\begin{align}
 \mathcal{M}_{ij} = \pdv{V}{\vb*{r}_i}{\vb*{r}_j} \quad i,j \notin P.
\end{align}
The eigenvalue problem of $\mathcal{M}$ is solved numerically using the Eigen package~\cite{eigenweb} to obtain all eigenvalues $\lambda_k$ and eigenvectors $\vb*{e}_k$.
For the system with $N = \num{40000}$, we use the SciPy package~\cite{Virtanen2020} to obtain the smallest eigenvalue and the corresponding eigenvector of this sparse matrix.

\section{Vibrational Density of States}
The vibrational density of states (VDOS) is
\begin{align}
 g(\omega) = \frac{1}{N_\text{mode}} \sum_k \delta\pqty{\omega - \omega_k},
\end{align}
where $N_\text{mode}$ is the number of all nonzero eigenmodes, and $\delta(x)$ is the Dirac delta function.
The number of configurations used to calculate the VDOS is given in Table~\ref{table:number}.

\begin{table}[H]
 \centering
 \caption{Number of samples used to calculate the VDOS.}
 \label{table:number}
 \begin{tabular}{rrrrrrrr}
  \toprule
         $c$ & 0.00        & 0.03        & 0.05        & 0.08        & 0.10        & 0.12        & 0.20 \\
  \midrule
  $N = 1000$ & \num{58800} & -           & \num{58800} & \num{58800} & \num{58800} & \num{58800} & \num{58800} \\
  $N = 4000$ & \num{24080} & \num{24080} & \num{24080} & \num{24080} & \num{24080} & \num{24080} & \num{24080} \\
  \bottomrule
 \end{tabular}
\end{table}

\begin{figure}[H]
\begin{minipage}{.49\textwidth}
 \centering
\includegraphics[width=\linewidth]{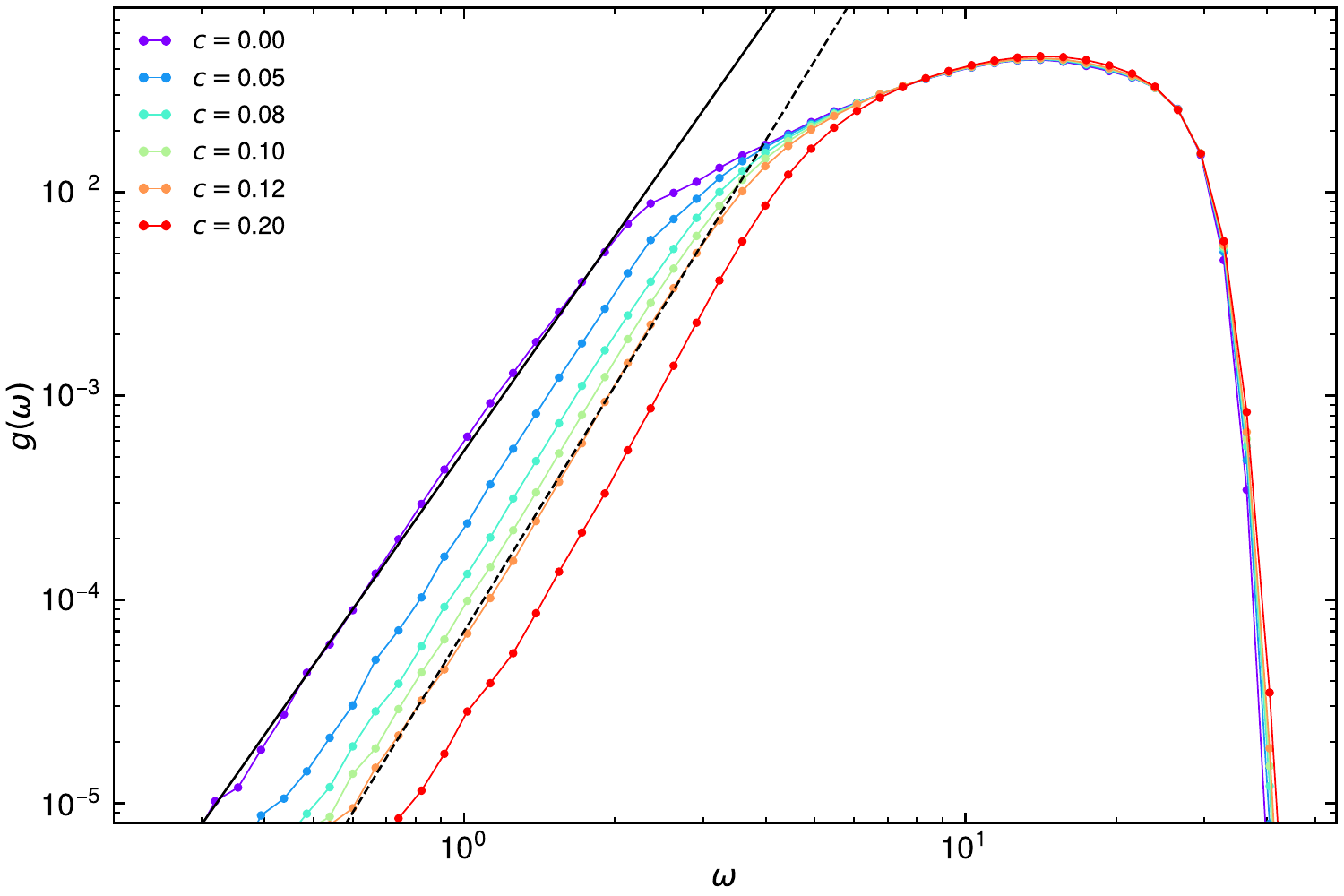}
\caption{The vibrational density of states for $N = 1000$. Solid and dashed lines indicate $g(\omega) \propto \omega^{3.5}$ and $g(\omega) \propto \omega^4$, respectively.}
 \label{fig:VDOS_KA1K}
\end{minipage}
 \hfill
\begin{minipage}{.49\textwidth}
 \centering
\includegraphics[width=\linewidth]{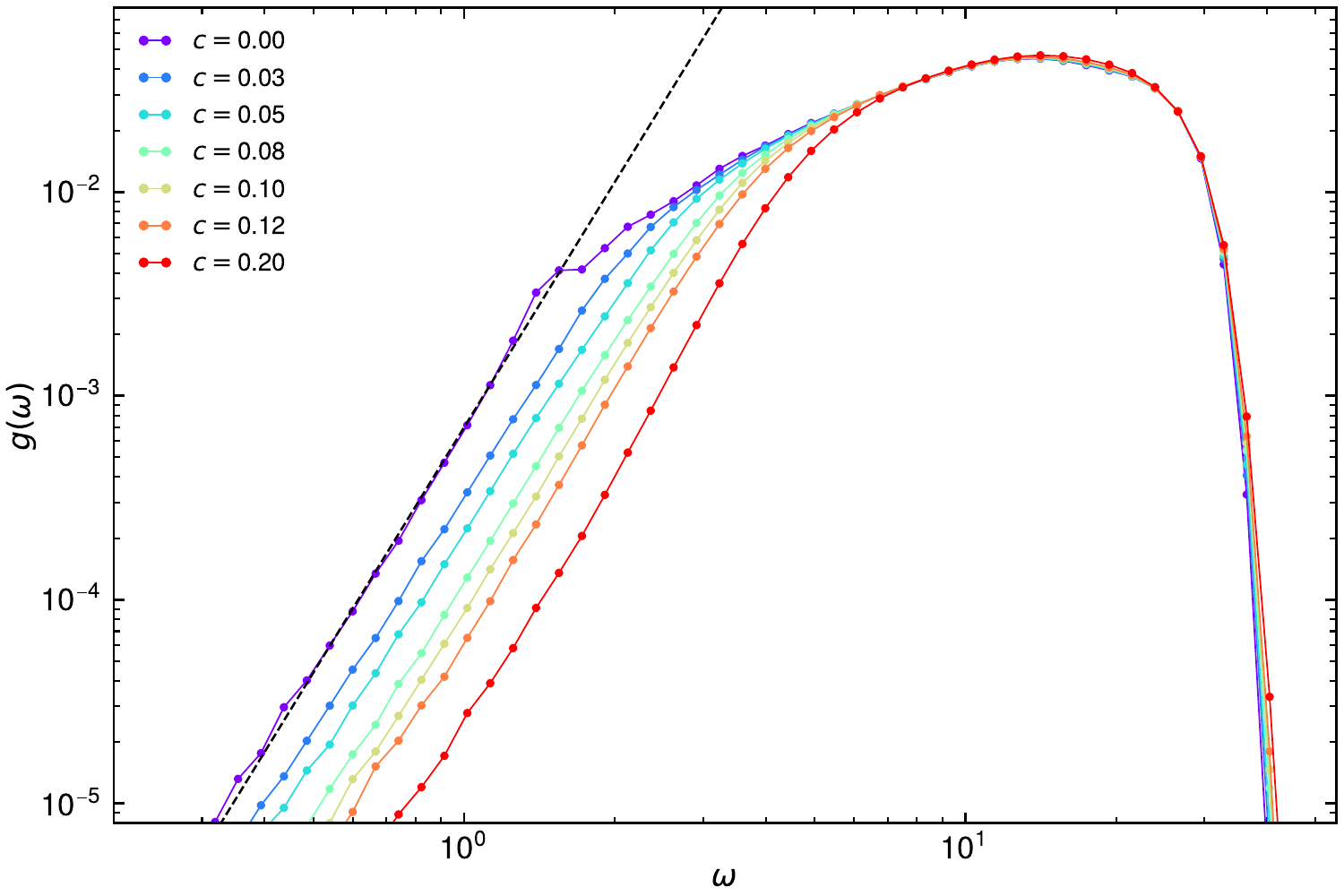}
 \caption{The vibrational density of states for $N = 4000$. Dashed line indicates $g(\omega) \propto \omega^4$.}
 \label{fig:VDOS_KA4K}
\end{minipage}%
\end{figure}

We calculated the VDOS for the cases of $N = 1000$ and $N = 4000$, as shown in Figs.~\ref{fig:VDOS_KA1K} and \ref{fig:VDOS_KA4K}.
For both cases, the VDOS with $c > 0$ shows $g(\omega) \propto \omega^4$ dependence at low frequency.
For the cases with $c = 0$, the VDOS of $N = 4000$ clearly shows the $\omega^4$ dependence, while the VDOS of $N = 1000$ shows $g(\omega) \propto \omega^{3.5}$.

